\documentclass[journal]{IEEEtran}

\ifCLASSINFOpdf
\else
\fi

\usepackage{array}
\usepackage{times}
\usepackage{epsfig}
\usepackage{graphicx}
\usepackage{amsmath}
\usepackage{amssymb}
\usepackage{graphicx}
\usepackage{cite}
\usepackage{enumerate}
\usepackage{cases}
\usepackage{multirow}
\usepackage{verbatim}
\usepackage{amssymb}
\usepackage{CJK}
\usepackage{subfigure}
\usepackage{algorithm}
\usepackage{algorithmicx}
\usepackage{algpseudocode}
\usepackage{stfloats}
\usepackage{color}
\usepackage{soul}
\usepackage{diagbox}
\usepackage{colortbl}
\usepackage{stfloats}

\usepackage{bm}
\usepackage{booktabs}
\usepackage[colorlinks,bookmarksopen,bookmarksnumbered,citecolor=black,linkcolor=black]{hyperref}

\begin{document}

\title{Diameter Estimation of Cylindrical Metal Bar Using Wideband Dual-Polarized Ground-Penetrating Radar}

\author{Hai-Han Sun, Weixia Cheng, and Zheng Fan

}

\maketitle

\begin{abstract}
Ground-penetrating radar (GPR) has been an effective technology for locating metal bars in civil engineering structures. However, the accurate sizing of subsurface metal bars of small diameters remains a challenging problem for the existing reflection pattern-based method due to the limited resolution of GPR. To address the issue, we propose a reflection power-based method by exploring the relationship between the bar diameter and the maximum power of the bar reflected signal obtained by a wideband dual-polarized GPR, which circumvents the resolution limit of the existing pattern-based method. In the proposed method, the theoretical relationship between the bar diameter and the power ratio of the bar reflected signals acquired by perpendicular and parallel polarized antennas is established via the inherent scattering width of the metal bar and the wideband spectrum of the bar reflected signal. Based on the theoretical relationship, the bar diameter can be estimated using the obtained power ratio in a GPR survey. Simulations and experiments have been conducted with different GPR frequency spectra, subsurface mediums, and metal bars of various diameters and depths to demonstrate the efficacy of the method. Experimental results show that the method achieves high sizing accuracy with errors of less than 10\% in different scenarios. With its simple operation and high accuracy, the method can be implemented in real-time in situ examination of subsurface metal bars.\par
\end{abstract}

\begin{IEEEkeywords}
Cylindrical metal bar, diameter estimation, dual polarization, ground-penetrating radar, scattering width, wideband.
\end{IEEEkeywords}

\IEEEpeerreviewmaketitle

\section{Introduction}

\IEEEPARstart {C}{ylindrical} metal bars are widely used in civil infrastructure, buildings, and utilities. Accurate measurement of the location and size of subsurface metal bars is important for assessing the health state of reinforced concrete structures \cite{importance3,importance4,importance1}. Although many nondestructive testing (NDT) methods have been developed with high effectiveness in locating the subsurface metal bar, the accurate sizing of the metal bar remains a challenging issue to be addressed \cite{gpr4concrete2, emi2}. \par

Ground-penetrating radar (GPR) and electromagnetic induction (EMI) sensors are two commonly used NDT technologies for inspecting subsurface metal bars \cite{gpr4concrete1,gpr4concrete2,emi2}. An EMI sensor receives the induced magnetic field response of a metal bar to the transmitted time-varying magnetic fields and establishes the relationship between the field strength and the bar diameter and cover depth by pre-calibrating a wide range of metal bars of different properties \cite{DL5}. Generally, EMI sensors such as cover meters require prior knowledge of the cover depth to accurately estimate the bar diameter \cite{emi1}. However, the precise cover depth is not always available in real scenarios, leading to unreliable diameter estimation results. Moreover, EMI sensors can only maintain their effectiveness within a short depth range of around 50 mm \cite{emi2,emi3}. 

GPR for inspecting subsurface metal bars usually employ antennas with their polarization parallel to the bar axis as a metal bar scatters maximally under this condition.  The GPR scans along a trace that is perpendicular to the axis of the metal bar to obtain the scattered waves from the metal bar at different spatial positions. The bar reflection is shown as a characteristic hyperbola in the B-scan radargram, and the geometry and position of the hyperbolic signature are dominated by the location and diameter of the metal bar and the relative permittivity of the surrounding medium. Reflection pattern-based methods including curve-fitting techniques \cite{CF1,CF2, CF3, CF4, CF5, curvefittinglimit} and generalized Hough transform algorithms \cite{HT1,HT2,HT3} have been developed to derive these parameters from the characteristic hyperbola. They have shown good accuracy in sizing cylindrical bars with diameters greater than 20 mm. However, their performance degrades for small-diameter bars as the limited resolution of the GPR cannot produce noticeable differences in the hyperbolic curvature with small variations in diameter\cite{curvefittinglimit,DL1}. Full-waveform inversion has been implemented to improve the sizing accuracy for small-diameter cases \cite{FWI1,FWI2,FWI3}. By incorporating information on the source wavelet in the parameter estimation process, it achieved higher estimation accuracy than conventional curve-fitting methods \cite{FWI3}. However, its iterative operation is computationally expensive, limiting its applicability in real-time in situ measurement.
Deep learning-based algorithms have been developed to enable automatic and real-time sizing of metal bars \cite{DL1,DL2,DL3,DL4}. The algorithms built a non-linear relationship between GPR data and metal bar parameters by training on a large labeled dataset. As a data-driven method, their applicability and accuracy heavily depend on the quality of the training data, yet building a large dataset covering diverse experimental scenarios is usually time-consuming and labour-intensive. \par

GPR systems equipped with antennas of different polarizations have been employed to improve detection accuracy and characterization of the metal bar and other elongated objects \cite{DL1,pol1,pol2,pol3,pol4,pol5,pol6,pol7}. The differential reflectivity (or power ratio) of orthogonal polarizations has been well used to estimate the median volume diameter of raindrops based on horizontal and vertical radar cross sections of oblate spheroids\cite{rain1, rain2, rain3}. Similarly, it has been experimentally demonstrated that the diameter of a thin metal bar can be estimated by the power ratio of bar reflected signals collected by antennas with polarizations perpendicular to and parallel to the bar axis \cite{DP1,DP2,DP3}. However, the method can only be used after taking measurements of many metal bars of known diameters in a given subsurface medium and generating a statistical power ratio distribution \cite{DP2}. Since the power ratio distribution varies with the frequency of the antenna and the subsurface medium, the applicability of this method in different experimental environments is limited \cite{DP1}. The scattering widths of the metal bar for orthogonally polarized electric fields have been used to explain the theory behind the method and to calculate the theoretical power ratio, but a large discrepancy occurs when comparing the power ratio measured by a GPR with the theoretical value calculated at the GPR's nominal  frequency point \cite{DP3}. This is because the GPR is a wideband device whose performance cannot be accurately characterized by a single frequency. Although the power ratio at a single frequency can be extracted and compared with the theoretical value, the reflected signal of a subsurface object at a single frequency point is more susceptible to random environmental noise, greatly reducing the stability of diameter estimation. Furthermore, the performance of the method with different GPR frequency spectra and subsurface mediums has not been fully investigated. In view of these problems and limitations, further research is needed to effectively use the wideband dual-polarized GPR to measure the size of a metal bar in different experimental environments. \par

This work devotes to addressing the aforementioned challenges of sizing metal bars using the wideband dual-polarized GPR. Firstly, the theoretical relationship between the bar diameter and the  power ratio of the bar reflected signal acquired by  wideband orthogonally polarized antennas is established by taking account of both the scattering width of the metal bar and the wideband spectrum of the reflected signal. Secondly, the validity of the theoretical relationship under different scenarios including the depth of the metal bar, the GPR operating frequencies, and the subsurface medium is investigated. Simulated and measured results demonstrate that the bar diameter can be reliably extracted from the acquired power ratio of the two orthogonal polarizations based on the established theoretical relationship. The proposed method achieves consistent estimation accuracy of the bar diameter in different experimental scenarios with an absolute percentage error of less than 10\%. \par

\section{Methodology}\label{sec2}

\subsection{Theoretical Foundation}

\begin{figure}[!t]
    \begin{center}
        \subfigure[]{
        \label{theo_per}
        \includegraphics[width=0.45\linewidth]{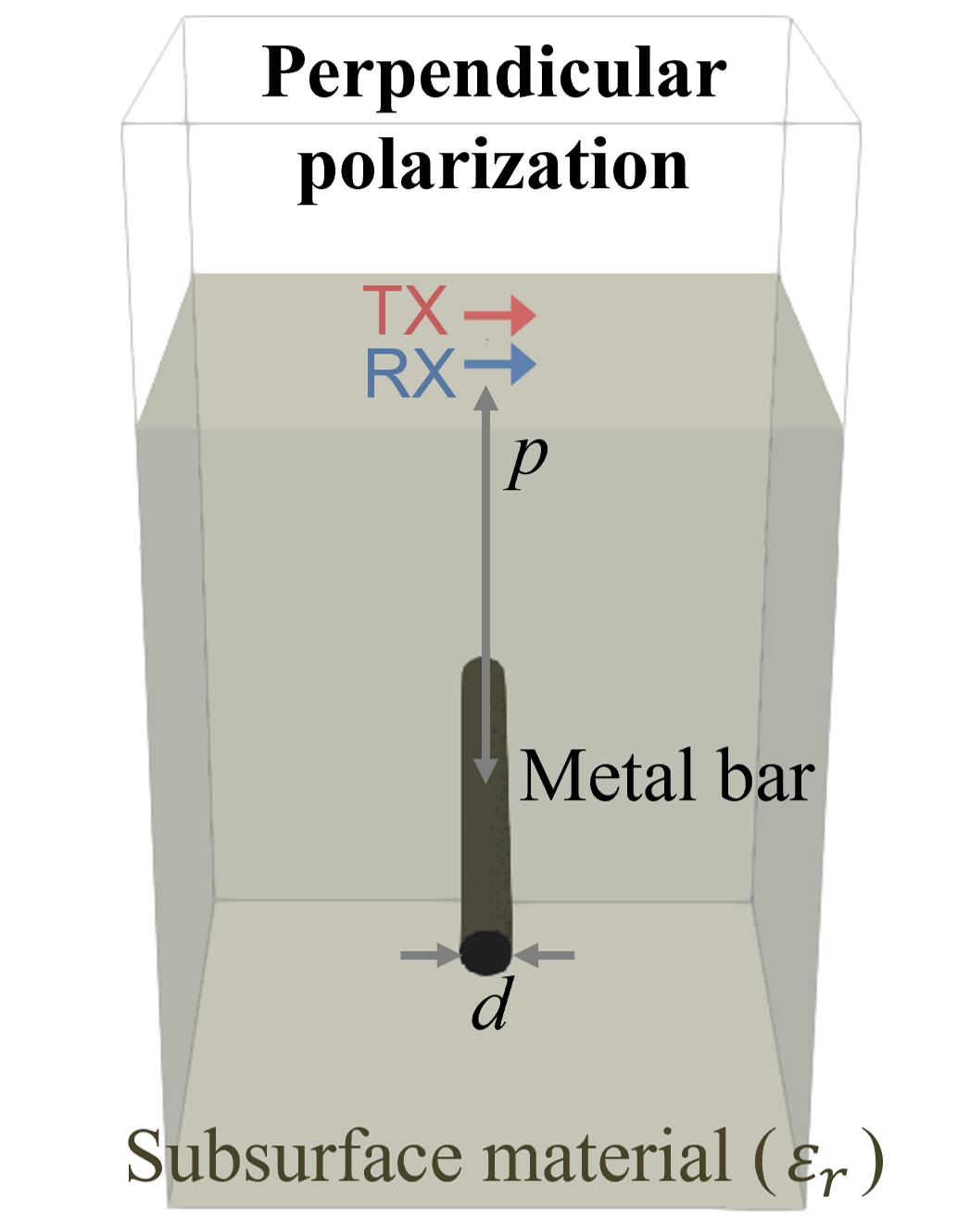}}
        \subfigure[]{
        \label{theo_para}
        \includegraphics[width=0.45\linewidth]{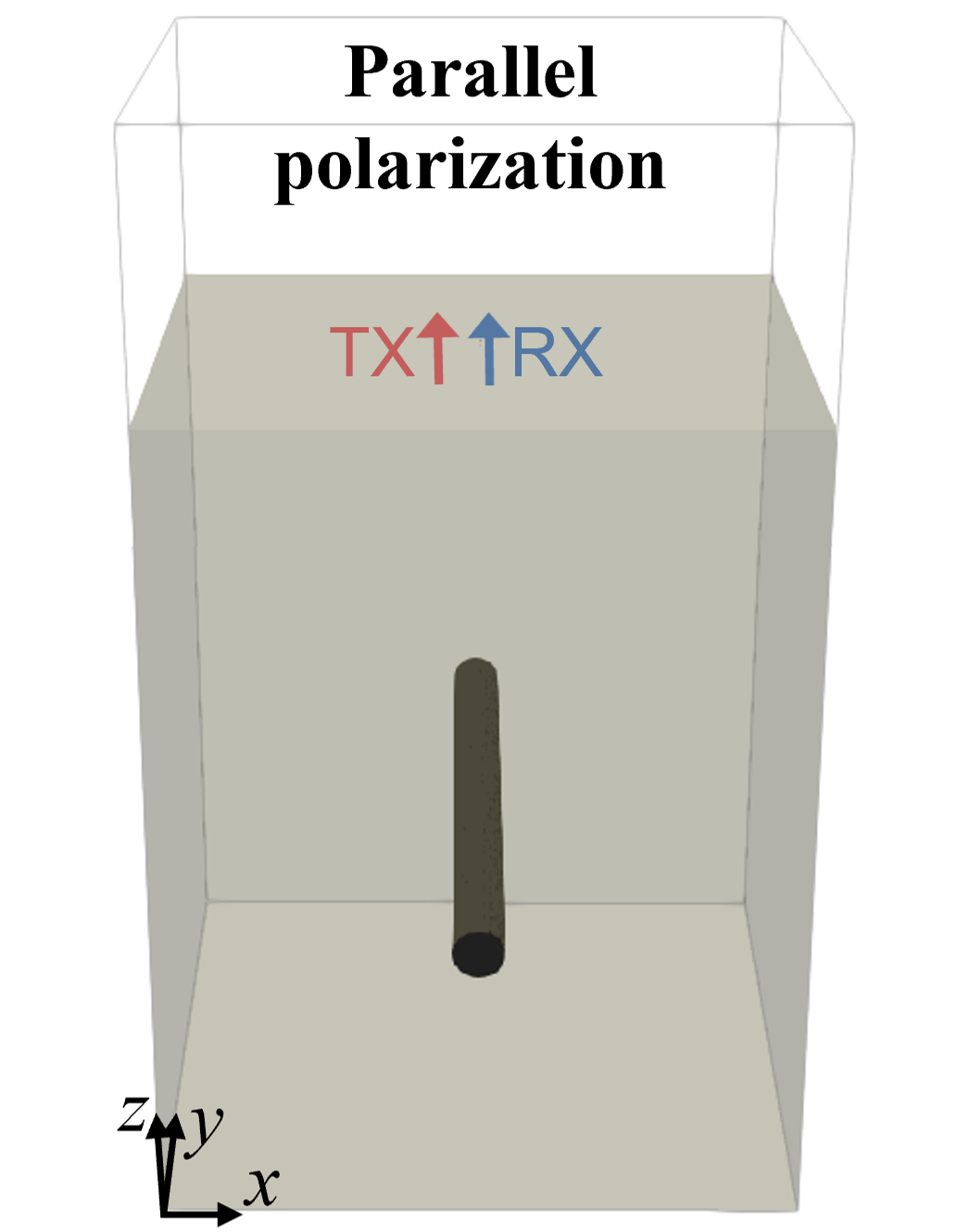}}
        \vspace{-0.1cm} 
        \caption{Illustration of the measurement scenario of using the transmitting antenna (TX) and receiving antenna (RX)  with (a) perpendicular polarization and (b) parallel polarization to detect a long cylindrical metal bar. The perpendicular and parallel directions are relative to the bar axis. $p$ and $d$ denote the depth and diameter of the metal bar, respectively.}
        \label{fig:theo_illustration}
    \end{center}
    \vspace{-0.5cm}
\end{figure}

In the proposed method, antenna systems with perpendicular and parallel polarizations relative to the bar axis are used to detect a long metal bar in a medium, as shown in Fig. \ref{fig:theo_illustration}. The transmitting and receiving antennas (TX and RX) can be the same antenna operating in a monostatic mode, or two antennas located adjacent to each other in a quasi-monostatic mode. The antennas with the parallel polarization and the perpendicular polarization have identical frequency responses and are excited with the same transmitting power. In practice, this can be achieved by using a single-polarized antenna and rotating it by 90$^{\circ}$ to realize the orthogonal polarizations, or using a specially designed dual-polarized antenna with same frequency characteristics for the two polarizations. \par

When the antennas are directly above the metal bar in the far field, the received power $P_r$  of the backscattered signal from the metal bar is calculated by the radar equation \cite{radarbook}: 
\begin{equation}
\label{eq_1}
P_r=\frac{P_t G_t}{L_t}\frac{1}{4\pi p^2L_{mt}}\sigma\frac{1}{4\pi p^2L_{mr}} \frac{G_r}{4\pi L_r} \left( \frac{c}{f\sqrt{\varepsilon_r}} \right)^2,
\end{equation}
where $P_t$  represents the transmitter power; $G_t$ and $G_r$ are the gain of the transmitting antenna and receiving antenna in the direction of the metal bar; $L_t$ and $L_r$ are losses in the transmitting and receiving systems; $L_{mt}$ and $L_{mr}$ are losses due to the propagating medium; $p$ is the distance between the TX/RX and the metal bar, $c$  is the speed of the light, $\epsilon_r$ is the relative permittivity of the surrounding medium, and $\sigma$ is the scattering width (or the scattering cross section per unit length) of the metal bar. Although the received power is dependent on many factors related to the GPR system and the propagation path, most of the factors can be eliminated by introducing the ratio of the signal power received by the perpendicular polarized antenna system $P_{r\perp}$ [Fig. \ref{theo_per}] to that received by the parallel polarized antenna system $P_{r\parallel}$ [Fig. \ref{theo_para}], producing     
\begin{equation}
\label{eq_2}
\frac{P_{r\perp}}{P_{r\parallel} } = \frac{\sigma_\perp}{\sigma_\parallel}.
\end{equation}
In the backscattering scenario as shown in Fig. \ref{fig:theo_illustration}, the scattering widths of the metal bar for the perpendicular and parallel polarized electric fields at frequency $f_0$ under the far-field condition ($p\gg \lambda_0$, and $\beta p\gg (\beta a)^2$) \cite{radarbook} are calculated by
\begin{equation}
\label{eq_3}
\sigma_{\perp}=\frac{4}{\beta}\left|\sum_{n=0}^{\infty}(-1)^{n+1} \zeta_{n} \frac{J_{n}^{\prime}(\beta a)}{H_{n}^{(1) \prime}(\beta a)}\right|^{2},
\end{equation}
and
\begin{equation}
\label{eq_4}
\sigma_{\|}=\frac{4}{\beta}\left|\sum_{n=0}^{\infty}(-1)^{n+1} \zeta_{n} \frac{J_{n}(\beta a)}{H_{n}^{(1)}(\beta a)}\right|^{2},
\end{equation}
where
\begin{equation}
\label{eq_5}
\begin{split}
&\beta=\frac{2 \pi f_0 \sqrt{\varepsilon_{r}}}{c}, \\
&\zeta_{n}= \begin{cases}1 & \text { for } n=0 \\ 2 & \text { for } n=1,2,3, \ldots\end{cases},
\end{split}
\end{equation}
$a$ is the radius of the metal bar, $\lambda_0$ is the wavelength,  $J_{n}(\cdot)$ is the usual cylindrical Bessel function of order n, $H_{n}^{(1)}(\cdot)$ is a cylindrical Hankel function of the first kind of order $n$. As shown in \eqref{eq_3} and \eqref{eq_4},  the scattering widths are directly related to the radius of the metal bar. Equations \eqref{eq_2}-\eqref{eq_5} allow us to build the theoretical relationship between the radius of the metal bar and the power ratio of reflected signals of the metal bar received by orthogonally polarized antennas at frequency $f_0$. \par

However, as the GPR is intrinsically a wideband device to guarantee high detection resolution and accuracy, instead of using a single frequency point, it is more reliable and convenient to calculate the ratio of the maximum power reflected by the metal bar and received by the two polarized antenna systems in the time domain \cite{DP1,DP2,DP3}. As the maximum received power of the reflected signal is contributed by the power at all frequency samples within the GPR operating band, the maximum power ratio $P_{r\perp max} /P_{r\| max}$ can be named as the wideband power ratio.  \par

Based on the inverse Fourier transform, the maximum amplitude of the received signal from the metal bar  $A_{r\, max}$ in the time domain is calculated by 
\begin{equation}
\label{eq_6}
A_{r \max }=\left|\frac{1}{N} \sum_{k=0}^{N-1} X_{r}(f_k) e^{j \frac{2 \pi s_{t} k}{N}}\right|,
\end{equation}
where $X_r(f_k)$ is the complex number that includes both amplitude and phase of signal at  the frequency point $f_k$, $k$ is the number of a frequency sample, $N$ is the total number of frequency samples within the GPR operating spectrum, and $s_{t}$ is the number of the time sample corresponding to the maximum amplitude point. The wideband power ratio $P_{r\perp max} /P_{r\| max}$ is therefore obtained by
\begin{equation}
\label{eq_7}
\frac{P_{r \perp max}}{P_{r \| max}}=\left(\frac{A_{r \perp max}}{A_{r \| max}}\right)^{2}=\left(\frac{\left|\sum_{k=0}^{N-1} X_{r \perp}\left(f_k\right) e^{j \frac{2 \pi s_{t} k}{N}}\right|}{\left|\sum_{k=0}^{N-1} X_{r \|}\left(f_k\right) e^{j \frac{2 \pi s_{t} k}{N}}\right|}\right)^{2}.
\end{equation}
Based on \eqref{eq_2}, the bar reflected signal received by the perpendicular polarized antenna and that received by the parallel polarized antenna at the frequency point $f_k$ satisfies
\begin{equation}
\label{eq_8}
\frac{X_{r \perp}\left(f_k\right)}{X_{r \|}\left(f_k\right)}= \sqrt{\frac{\sigma_{\perp}\left(f_k\right)}{\sigma_{\|}\left(f_k\right)}} .
\end{equation}
Substituting \eqref{eq_8} into \eqref{eq_7} produces 
\begin{equation}
\label{eq_9}
\frac{P_{r \perp max}}{P_{r \| max}}=\left(\frac{\left| \sum_{k=0}^{N-1} X_{r \|}\left(f_k\right) \sqrt{\frac{\sigma_{\perp}\left(f_k\right)}{\sigma_{ \|}\left(f_k\right)}}  e^{j \frac{2 \pi s_{t} k}{N}}\right|}{\left|\sum_{k=0}^{N-1} X_{r \|}\left(f_{k}\right) e^{j \frac{2 \pi s_{t} k}{N}}\right|}\right)^{2} .
\end{equation}
The full equation of \eqref{eq_9} after substituting \eqref{eq_3} and \eqref{eq_4} is
\begin{equation}
\label{eq_10}
\begin{aligned}
&\frac{P_{r \perp max}}{P_{r \| max}}= \\ 
& \left(\frac{\left|\sum_{k=0}^{N-1} X_{r \|}\left(f_{k}\right) \frac{\left|\sum_{n=0}^{\infty}(-1)^{n+1} \zeta_{n} \frac{J_{n}^{\prime}(\beta a)}{H_{n}^{(1) \prime}(\beta a)}\right|}{\left|\sum_{n=0}^{\infty}(-1)^{n+1} \zeta_{n} \frac{J_{n}(\beta a)}{H_{n}^{(1)}(\beta a)}\right|}e^{j \frac{2 \pi s_{t} k}{N}}\right|}{\left|\sum_{k=0}^{N-1} X_{r \|}\left(f_{k}\right)e^{j \frac{2 \pi s_{t} k}{N}}\right|}\right)^{2}.
\end{aligned}
\end{equation}
Equation \eqref{eq_10} builds the theoretical relationship between the wideband power ratio $P_{r\perp max} /P_{r\| max}$ and the bar radius $a$. 

\subsection{Implementation Process} \label{procedure}
Based on \eqref{eq_10}, the bar size can be estimated using dual-polarized GPR data following three steps. 
\begin{enumerate} [1.]
\item	\textbf{Calculate the wideband power ratio of the acquired dual-polarized GPR data.} After obtaining the A-scan containing metal bar reflections using perpendicular and parallel polarized antenna systems, the metal bar reflections are first extracted by subtracting the background trace measured in the same environment but without the metal bar from the A-scan. The wideband power ratio $P_{r\perp max}/P_{r\| max} $ is then calculated as the square of peak amplitudes of the bar reflected signals in the time domain. 
\item	\textbf{Plot theoretical curve between the bar diameter and the wideband power ratio.} The theoretical relationship between bar diameter and the wideband power ratio is calculated using \eqref{eq_10}. The spectrum $X_{r\|}(f_k)$ in the equation is obtained by applying Fourier transform to the reflected signal of the metal bar acquired by the parallel polarized antenna in the time domain. $s_t$ is the number of the time sample corresponding to the maximum amplitude point. The relative permittivity of the subsurface medium $\epsilon_r$  needs to be pre-determined and substituted into the equation. There are several GPR methods to measure $\epsilon_r$ of an given medium, such as the delay time-velocity method using an object of a known depth \cite{vt} and the curve fitting method \cite{CF1,CF2,CF3,CF4,CF5}. Dielectric probes can also be used to directly measure the dielectric property of a medium \cite{probe1,probe2,probe3}, which are not described in detail here.  The theoretical curve for a list of diameters and the corresponding wideband power ratios can then be calculated and plotted. 
\item	\textbf{Estimate the bar diameter based on the theoretical curve.} The size of the metal bar can be extracted by finding the diameter value corresponding to the measured wideband power ratio in the theoretical curve. 
\end{enumerate}

\subsection{Case Study} 

\begin{figure}[!th]
    \begin{center}
        \subfigure[]{
        \label{realtoabs}
        \includegraphics[width=0.95\linewidth]{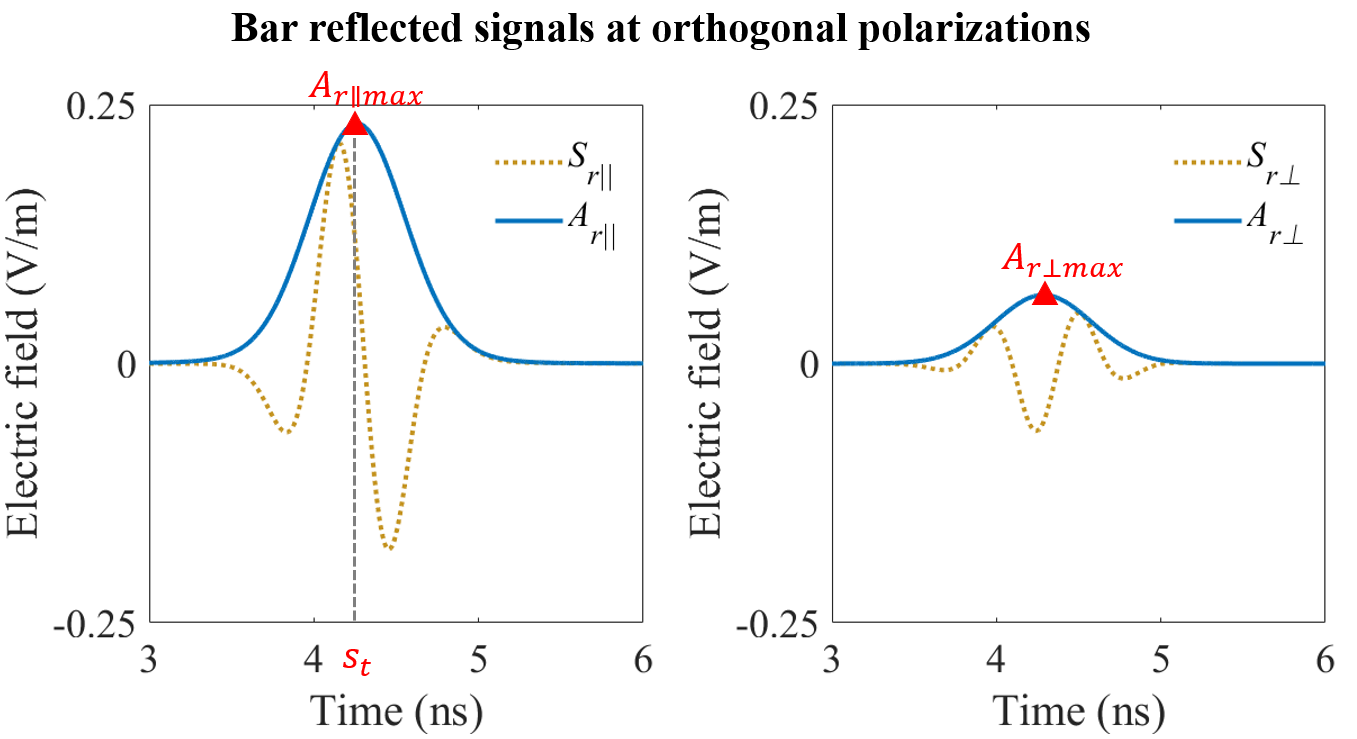}}\\
        \vspace{-0.1cm} 
        \subfigure[]{
        \label{frequencycomponent}
        \includegraphics[width=0.75\linewidth]{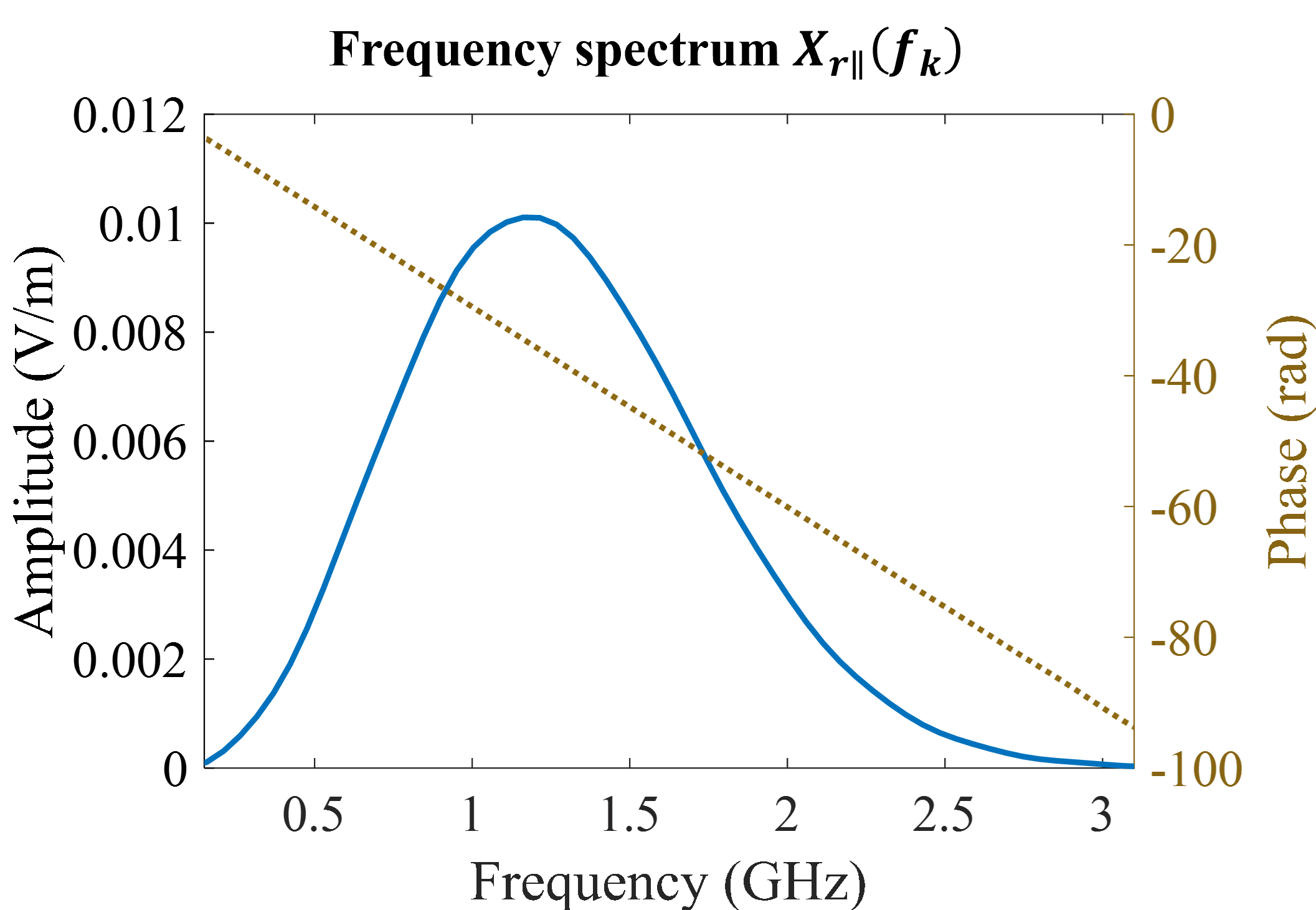}}\\
        \vspace{-0.1cm} 
        \subfigure[]{
        \label{determineeps}
        \includegraphics[width=0.82\linewidth]{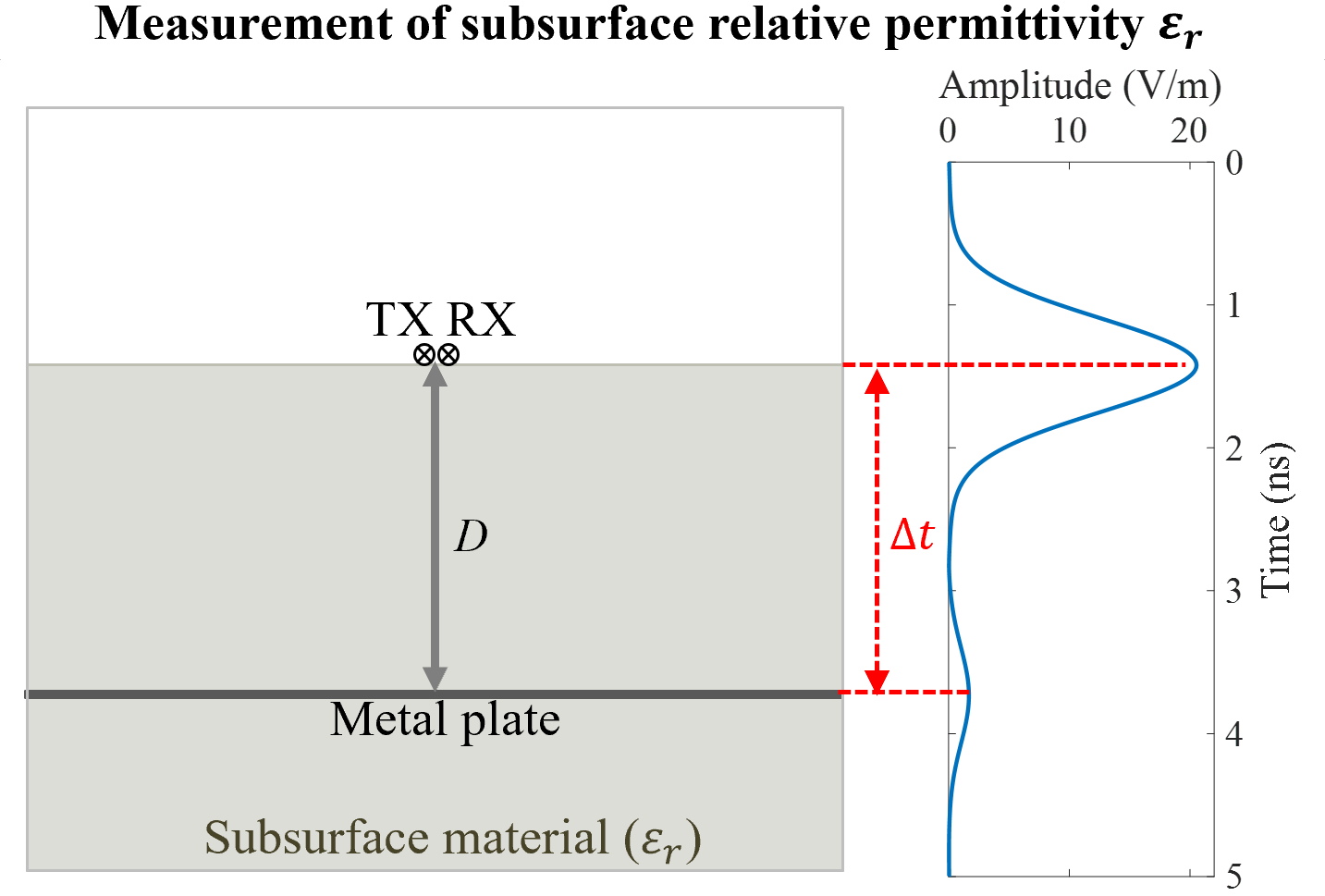}}\\
        \vspace{-0.1cm} 
        \subfigure[]{
        \label{theocurve}
        \includegraphics[width=0.77\linewidth]{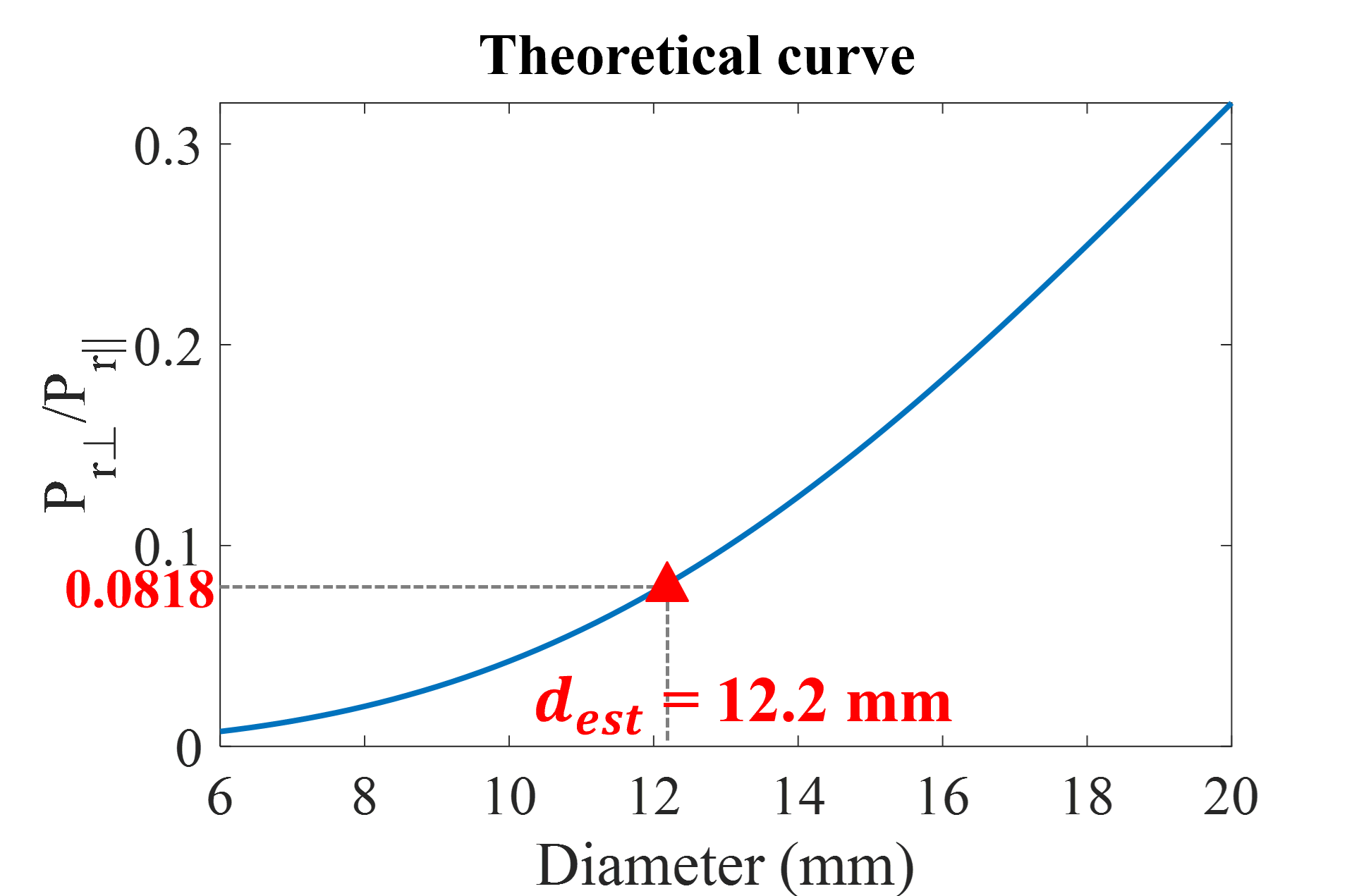}}\\
        \vspace{-0.1cm} 
        \caption{The results of a case study. (a) The bar reflected signals at the orthogonal polarizations. (b) The frequency spectrum of the bar reflected signal obtained using Fourier transform. (c) The determination of the subsurface relative permittivity. (d) The calculated theoretical curve by \eqref{eq_10}. The estimated diameter using the measured wideband power ratio and the theoretical curve is 12.2 mm, which is very close to the real diameter 12.0 mm.}
        \label{fig:demo}
    \end{center}
    \vspace{-0.5cm} 
\end{figure}

A case study is presented to illustrate the implementation process of the method. The scenario is shown in Fig. \ref{fig:theo_illustration}, where an impulse GPR is used to detect a metal bar with a diameter ($d$) of 12 mm at a depth ($p$) of 30 cm in a medium with relative permittivity ($\epsilon_r$) of 3. The source waveform is the Ricker waveform with 1 GHz center frequency. The data in the case study is acquired using the open-source software gprMax \cite{gprmax1,gprmax2}. Other simulation settings are the same as those described in the numerical studies in Section \ref{sec3}. \par
Following step 1, the extracted bar reflected signals acquired by the parallel and perpendicular polarized antennas ($S_{r \|}$ and $S_{r \perp}$) are shown in Fig. \ref{realtoabs}. Hilbert transform is used to convert the real-value signal to its complex representation to obtain its amplitude. The wideband power ratio is calculated as $\left(A_{r \perp \max } / A_{r \| \max }\right)^{2}=0.0818$. The number of time sample corresponding to the maximum amplitude point $s_t$ is also obtained from the figure. \par
Following step 2, the spectrum $X_{r\|}(f_k)$ is obtained by applying Fourier transform to the bar reflected signal acquired by the parallel polarized antenna, as shown in Fig. \ref{frequencycomponent}. The relative permittivity of the subsurface medium is pre-determined by placing a metal plate at a depth $D=20$ cm in the medium and measuring the delay time from the surface to the plate $\Delta t$, as illustrated in Fig. \ref{determineeps}. The relative permittivity is obtained by $\epsilon_r=(c\Delta t/2D)^2=3.0$. The wideband power ratios of a list of diameters are calculated by substituting $s_t$, $X_{r\|}(f_k)$, and $\epsilon_r$ into \eqref{eq_10}, and the resulting theoretical curve is shown in Fig. \ref{theocurve}. Given the measured wideband power ratio and the theoretical curve, the diameter is estimated to be 12.2 mm. The estimated value is very close to the real diameter 12.0 mm with a percentage error of 1.7\%, demonstrating the effectiveness of the method.  \par
Extensive simulations and experiments have been conducted to demonstrate the performance of the method in different scenarios, which are presented in Section \ref{sec3} and Section \ref{sec4}.

\section{Simulation Results}\label{sec3}

Numerical simulations are performed using the open-source software gprMax \cite{gprmax1,gprmax2}. The simulated model is built following the one illustrated in Fig. \ref{fig:theo_illustration}. The simulated domain covers an area of 0.3$\times$0.3$\times$0.7 $m^3$. The absorbing boundary condition is applied to reduce the boundary reflections. The hertzian dipole and the probe are used as the TX and RX, respectively. They are 10 mm apart and are located on the ground surface with their common middle point at 0.15 m along the $x$- and $y$-axes. Both the perpendicular polarization and the parallel polarization of the TX and RX are simulated, as shown in Figs. \ref{theo_per} and \ref{theo_para} respectively. The object is a cylindrical metal bar made of the perfect electric conductor (PEC). Its axis is parallel to the $y$-axis and is located at 0.15 m along the $x$-axis. \par

\begin{figure}[!t]
    \begin{center}
        \subfigure[]{
        \label{depth1}
        \includegraphics[width=0.9\linewidth]{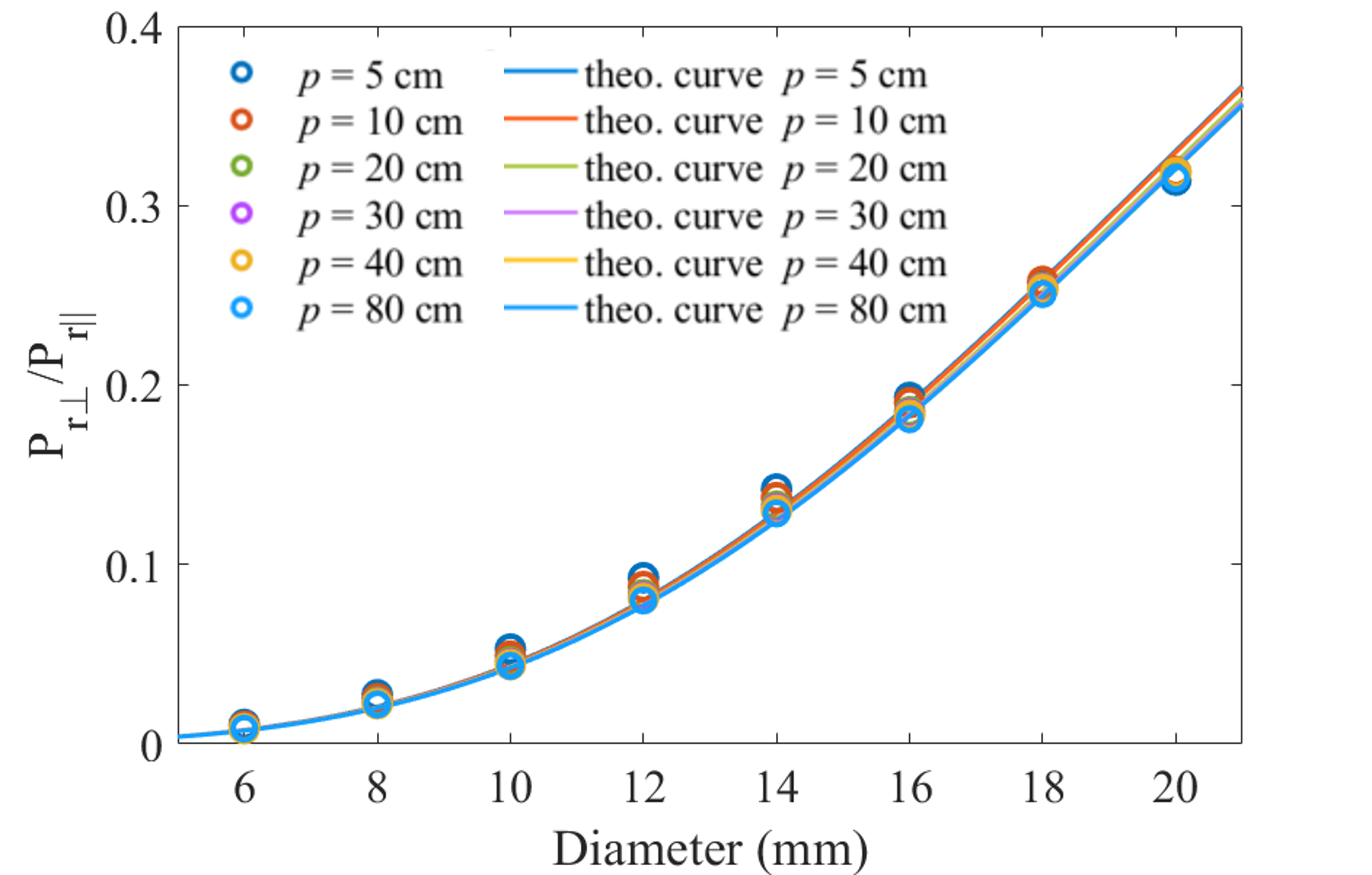}}\\
        \vspace{-0.1cm} 
        \subfigure[]{
        \label{depth2}
        \includegraphics[width=0.9\linewidth]{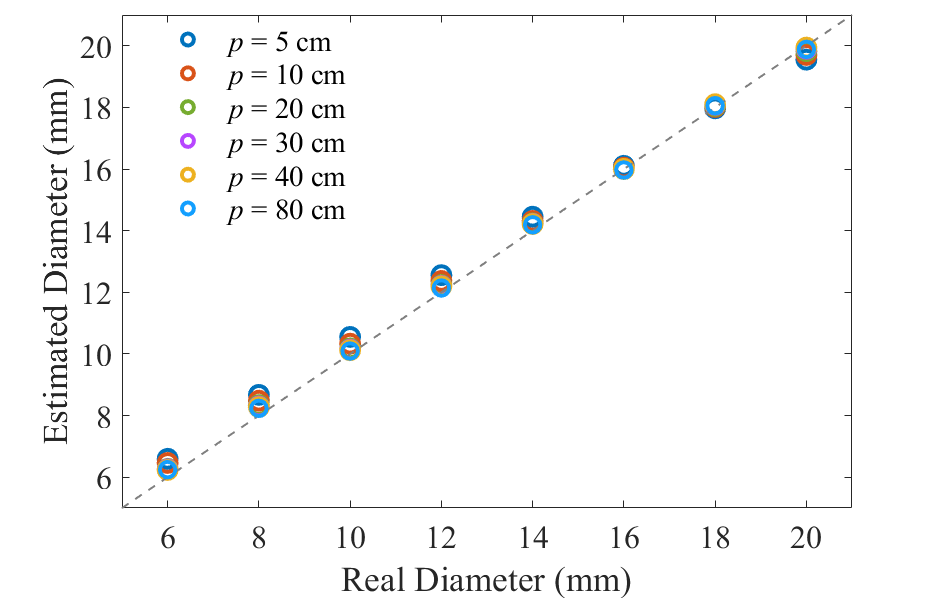}}\\
        \vspace{-0.1cm} 
        \subfigure[]{
        \label{depth3}
        \includegraphics[width=0.9\linewidth]{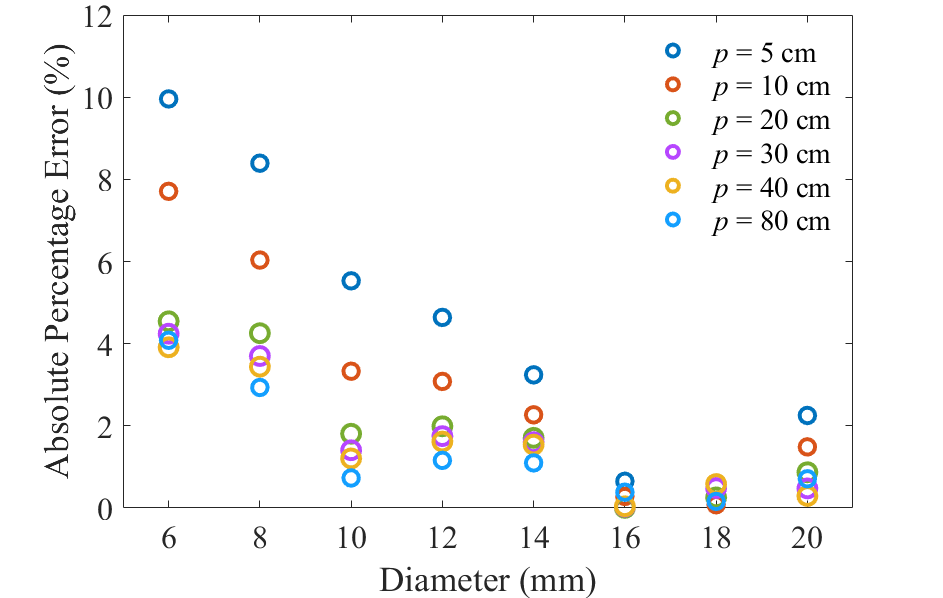}}\\
        \vspace{-0.1cm} 
        \caption{(a) The calculated power ratio compared with the theoretical curve for the metal bar of different diameters at different depths. (b) The estimated diameters using the proposed method. (c) The absolute percentage errors of the estimated diameters. }
        \label{fig:depth}
    \end{center}
    \vspace{-0.5cm}
\end{figure}

Since \eqref{eq_10} is derived under the far-field condition, and it shows that the wideband power ratio is related to the metal bar's diameter, the GPR frequency spectrum, and the medium relative permittivity, simulations are performed with a metal bar of different diameters at different depths, GPR transmitter of different frequency spectra, and the subsurface medium of different relative permittivity to investigate the effectiveness of the method. \par

\textbf{Depth.} In the simulation, a metal bar with its diameter $d$ ranging from 6 mm to 20 mm with a step of 2 mm is located at five different cover depths $p$ from 5 cm to 80 cm. The source waveform is the Ricker waveform with 1 GHz center frequency. The relative permittivity of the subsurface medium is set as 3. To guarantee good discretization for the metal bar of the smallest diameter, a fine spatial discretization step of 0.6 mm is used in the $x$-, $y$-, and $z$-directions. A-scans of the perpendicular and parallel polarizations are obtained for the metal bar of different diameters at different depths. \par

Following the implementation process presented in Sections II.B and II.C, the reflection of the metal bar is extracted by subtracting a background A-scan without the metal bar from the received A-scan, and then the wideband power ratio $P_{r\perp max}/P_{r \parallel max}$ is calculated. The wideband power ratio of the metal bar of different diameters at different depths is plotted in Fig. \ref{depth1}. It can be seen that the power ratio for the same diameter varies slightly in the shallow-depth cases ($p$ = 5 cm and $p$ = 10 cm), but tends to be constant for depths over 20 cm. According to the theory of the metal bar's scattering widths \cite{radarbook}, the far-field condition is satisfied when the distance between the antenna and the metal bar is greater than the operating wavelength at the center frequency $\lambda = c/f\sqrt{\epsilon_r}=17.32$ cm. Therefore, the variation of the power ratio in shallow-depth cases is because the distance is in the near field where the electromagnetic waves may not be fully polarized when illuminated on the metal bar, so the resultant backscattered power is different from that in the far field. Upon approaching the far field, the power ratio reaches a stable value regardless of the depth variation, which is consistent with the theoretical relationship described in \eqref{eq_10} that the wideband power ratio is independent of the depth. Nevertheless, the power ratio in the near field when the depth exceeds half wavelength does not deviate much from the value obtained in the far field. \par

\begin{figure}[!t]
    \begin{center}
        \subfigure[]{
        \label{ricker1}
        \includegraphics[width=0.9\linewidth]{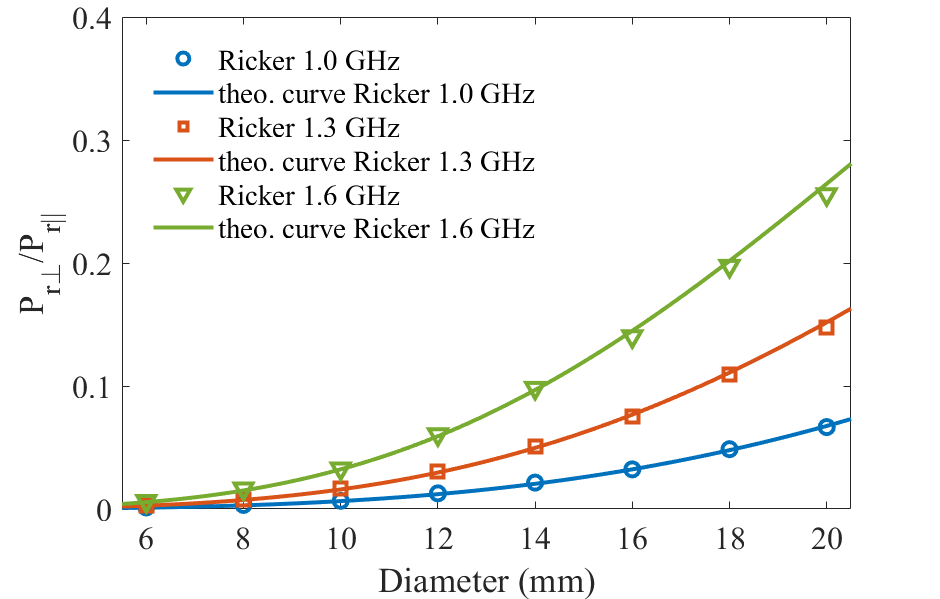}}\\
        \vspace{-0.1cm} 
        \subfigure[]{
        \label{ricker2}
        \includegraphics[width=0.9\linewidth]{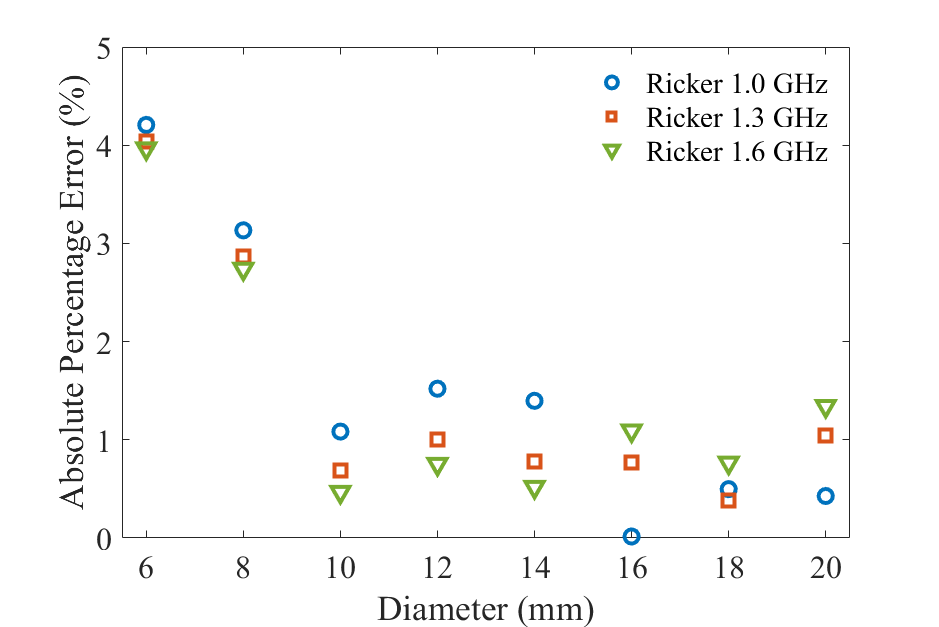}}\\
        \vspace{-0.1cm} 
        \caption{(a) The power ratio of the metal bar and the theoretical curves obtained using Ricker waveform with different center frequencies. (b) The absolute percentage errors of the estimated diameters.}
        \label{fig:ricker}
    \end{center}
    \vspace{-0.5cm}
\end{figure}

The theoretical curves depicting the relationship between the wideband power ratio and the bar size are calculated using \eqref{eq_10}. Since the received spectrum $X_{r\|}(f_k)$ of the metal bar of different diameters at the same depth is almost identical as found in the simulation, the spectrum used to calculate the curve at each depth is the averaged spectrum obtained from all diameters at that depth. The theoretical curves at different depths are also presented in Fig. \ref{depth1}. The curves overlap with each other, especially those for the distance over 20 cm, which further verifies that the relationship between the power ratio and the bar size is independent of the bar depth under the far-field condition. \par

As shown in Fig. \ref{depth1}, the simulated wideband power ratios at different diameters fit well with the theoretical curves. The estimated diameters using the wideband power ratios and the theoretical curves are shown in Fig. \ref{depth2}, and the absolute percentage errors are shown in Fig. \ref{depth3}. The estimated diameters are close to the real values. The 5 cm depth cases have the lowest accuracy due to the deviation of the wideband power ratio in the near field. The accuracy improves greatly as the depth increases. For the 10 cm depth cases, the errors are within 8\%. For depths over 20 cm, the accuracy is further improved to less than 5\% errors. The largest error at each depth occurs when the metal bar has the smallest diameter 6 mm. This is because the slope of the theoretical curve is very small around the 6 mm diameter region, so a small deviation of the power ratio due to the numerical simulation of finite spatial discretization can result in a relatively large error. The simulated results demonstrate that the method works the best under the far-field condition where the depth is larger than the wavelength, and maintains good accuracy of less than 10\% errors when the depth is over half wavelength in the near-field region. The 80-cm depth case verifies that the method maintains its high accuracy as long as GPR can successfully detect the reflected signal. The maximum detection range of the GPR depends on the system power, GPR spectrum, and subsurface environments, which varies from case to case. Similar phenomena are observed using different GPR operating frequencies and subsurface permittivity in our study. Therefore, we suggest to apply the method to scenarios where the distance between the GPR antenna and the metal bar exceeds half wavelength at the center frequency.  \par

\begin{figure}[!t]
    \begin{center}
        \subfigure[]{
        \label{permittivity1}
        \includegraphics[width=0.9\linewidth]{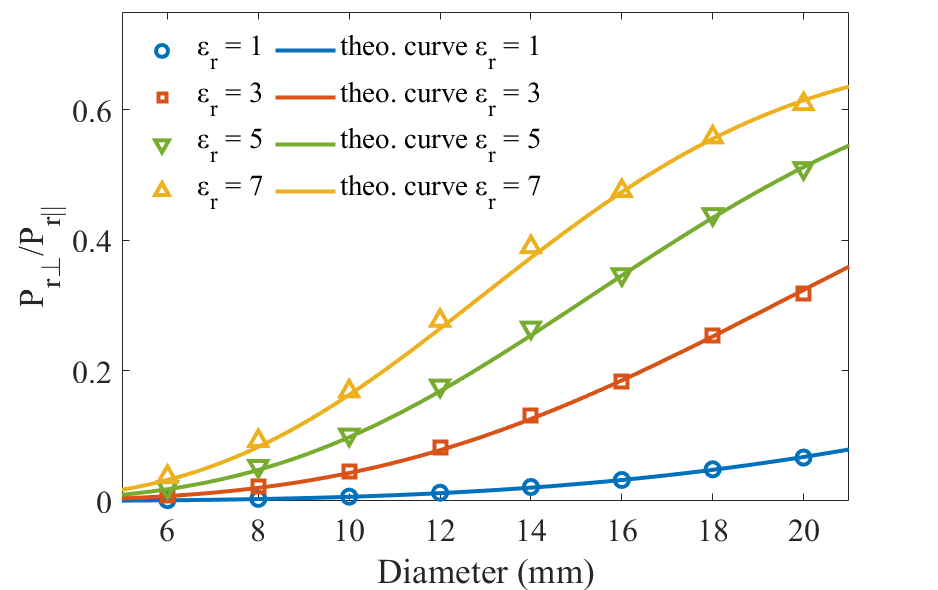}}\\
        \vspace{-0.1cm} 
        \subfigure[]{
        \label{permittivity2}
        \includegraphics[width=0.9\linewidth]{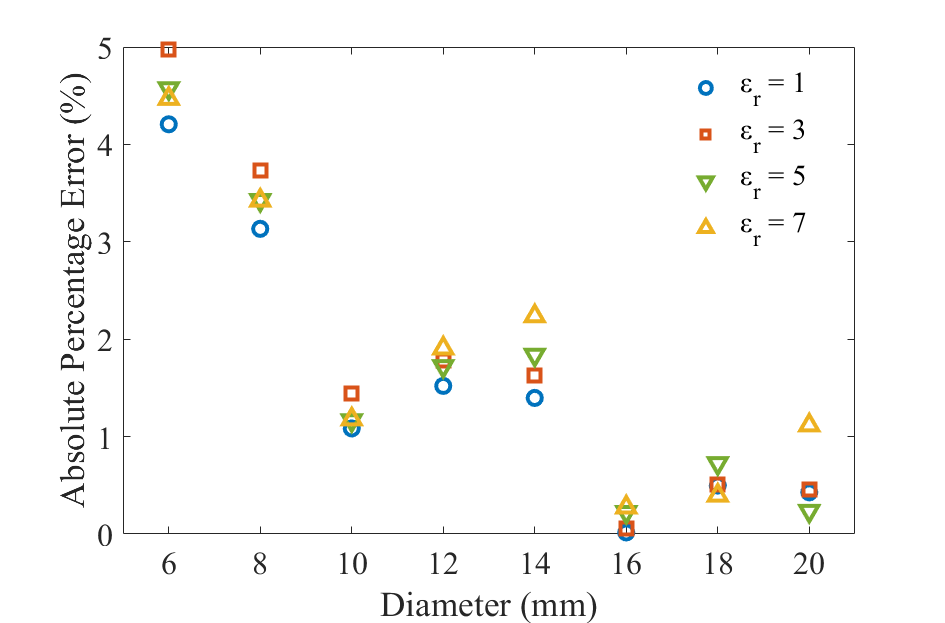}}\\
        \vspace{-0.1cm} 
        \caption{(a) The power ratio of the metal bar and the theoretical curves in a subsurface medium of different relative permittivity. (b) The absolute percentage error of the estimated diameters.}
        \label{fig:permittivity}
    \end{center}
    \vspace{-0.5cm}
\end{figure}

\textbf{Frequency spectrum.} To investigate the validity of the proposed method for GPRs with different operating frequency spectra, the Ricker waveform with three different center frequencies of 1.0 GHz, 1.3 GHz, and 1.6 GHz are used as the source waveform in the simulation. The relative permittivity of the medium is 1. The depth of the metal bar is 30 cm, which satisfies the far-field condition. The obtained wideband power ratio and the calculated theoretical curves with the different spectra are shown in Fig. \ref{ricker1}. As the center frequency of the waveform increases, the wideband power ratio for a metal bar of the same diameter also increases. The wideband power ratio is consistent with the corresponding theoretical curve in all cases. The estimated diameters based on the theoretical curve are very close to the real diameters. The maximum error is 4.25\% and most errors are less than 2.0 \%, as shown in Fig. \ref{ricker2}. The results demonstrate the effectiveness of the proposed method with different GPR operating spectra.

\begin{figure}[t]
	\centering
	\centerline{\includegraphics[width=1.0\linewidth]{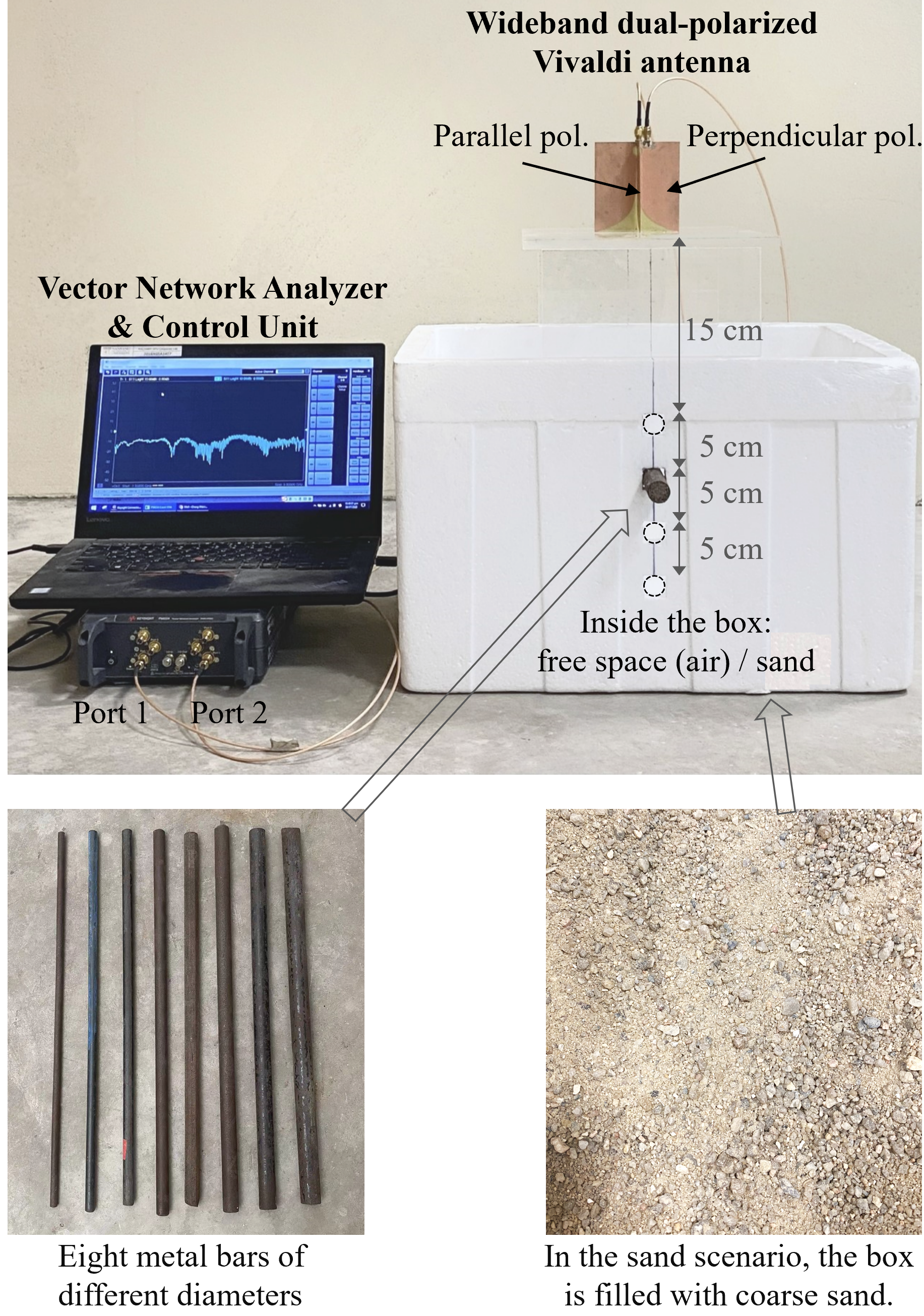}}
	\caption{Illustration of the measurement scenario in the air and in the coarse sand environment. A wideband dual-polarized GPR system is configured using a dual-polarized Vivaldi antenna and a vector network analyzer to obtain the reflected signals of the metal bar. }
	\label{fig:conf}
	\vspace{-0.5cm}
\end{figure}

\textbf{Relative permittivity of medium.} The subsurface medium with its relative permittivity values ranging from 1 to 7 with a step of 2 is simulated to examine the performance of the proposed method in different mediums. The source waveform is the Ricker waveform with a center frequency of 1.0 GHz. The cover depth of metal bars is 30 cm. The power ratio of the metal bar in different mediums is shown in Fig. \ref{permittivity1}. The calculated theoretical curves are also presented for comparison. As the relative permittivity increases, the electrical size of the metal bar becomes larger, leading to an increase in power ratio. The power ratios in different mediums are in good agreement with the corresponding theoretical curves. The absolute percentage error of the estimated diameters based on the theoretical curves is shown in Fig. \ref{permittivity2}. The errors are within 5\% in all cases, which verifies the effectiveness of the proposed method in mediums of different relative permittivity values.\par

\section{Measurement Results}\label{sec4}
Experiments are conducted with three different mediums, air, coarse sand, and concrete, to verify the performance of the proposed methods. A dual-polarized stepped-frequency GPR system is used in the experiment, as shown in Fig. \ref{fig:conf}. A dual-polarized Vivaldi antenna operating from 0.5 GHz to 3.3 GHz is used as the transmitter and receiver in a monostatic setup. The antenna has two ports to excite the perpendicular polarized and parallel polarized radiation with identical frequency responses. The two ports are connected to a vector network analyzer (VNA, Keysight VNA P5022A) that is used for the generation of the excitation signal and the acquisition of the reflected signal in the frequency domain. Other detailed setups and the experimental results in the three mediums are presented in the following subsections. \par

\begin{figure*}[!t]
    \begin{center}
        \subfigure[]{
        \label{air1}
        \includegraphics[width=0.33\linewidth]{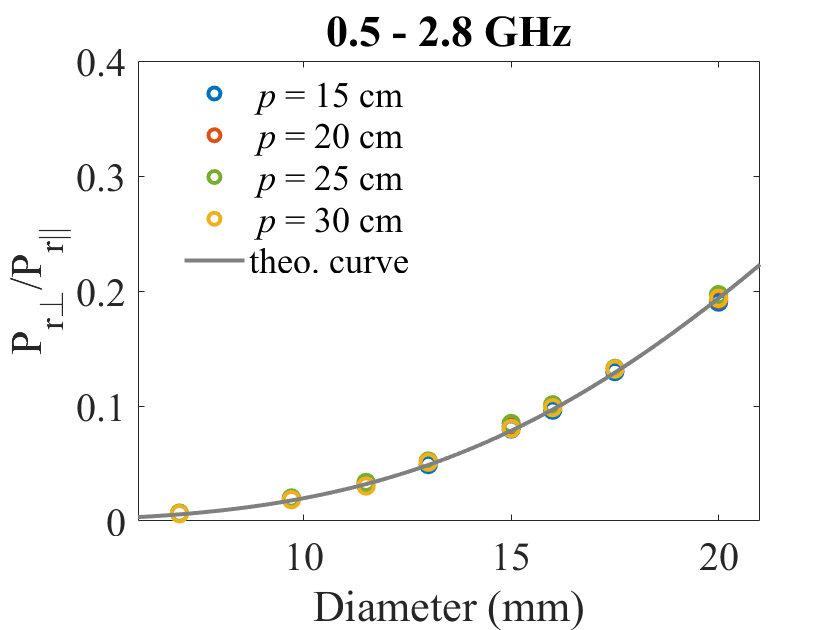} 
		\includegraphics[width=0.33\linewidth]{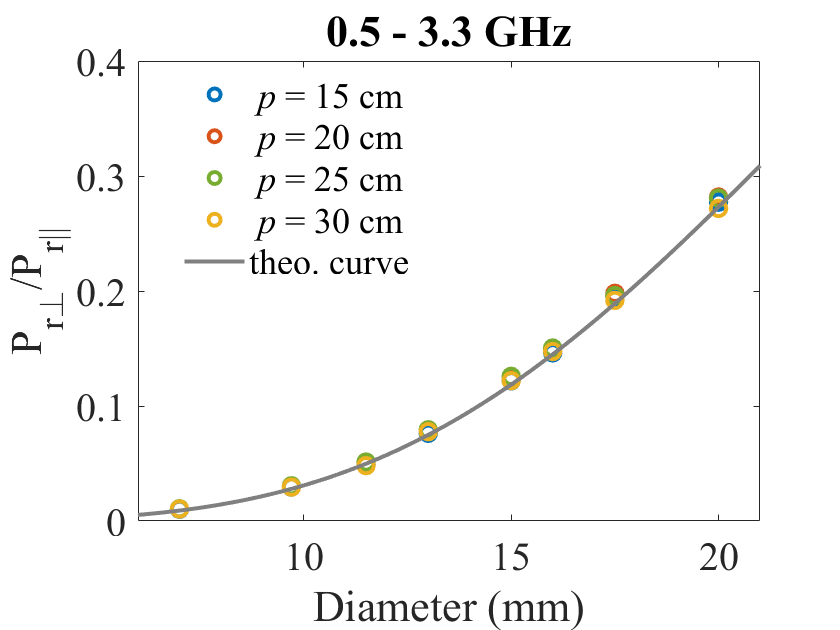} 
		\includegraphics[width=0.33\linewidth]{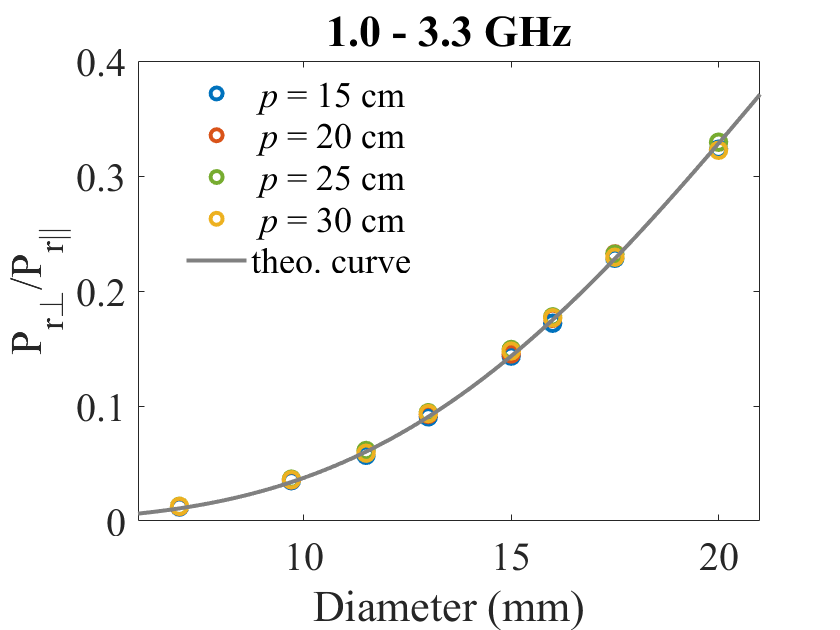}}\\
        \subfigure[]{
        \label{air2}
        \includegraphics[width=0.33\linewidth]{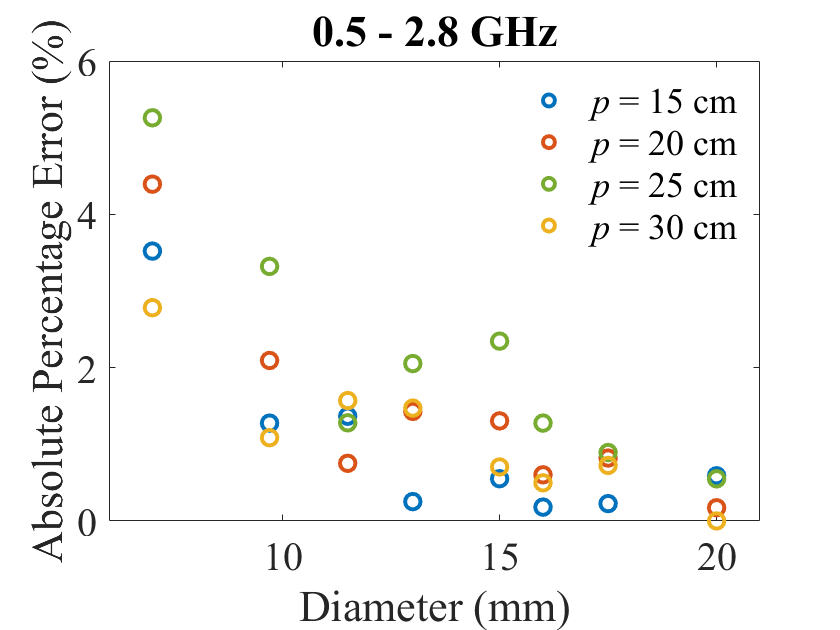} 
		\includegraphics[width=0.33\linewidth]{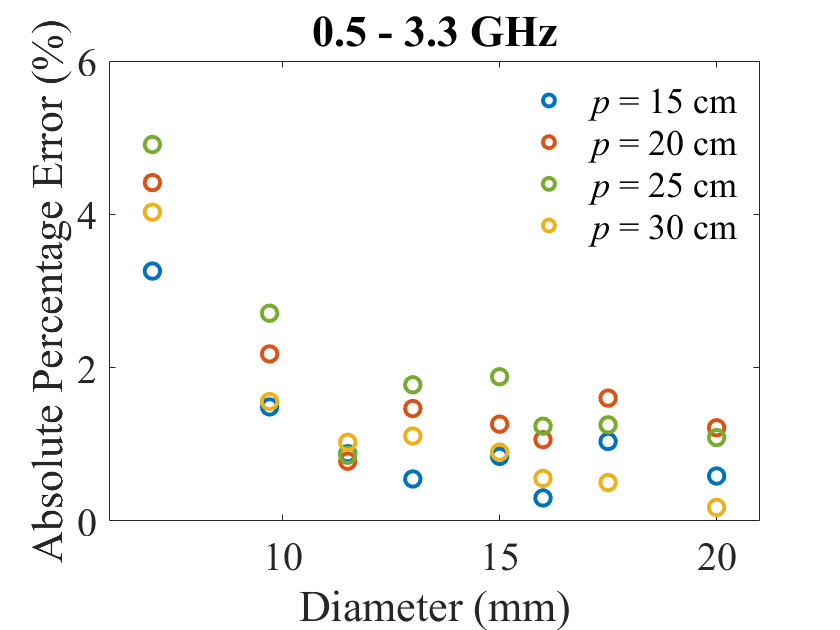} 
		\includegraphics[width=0.33\linewidth]{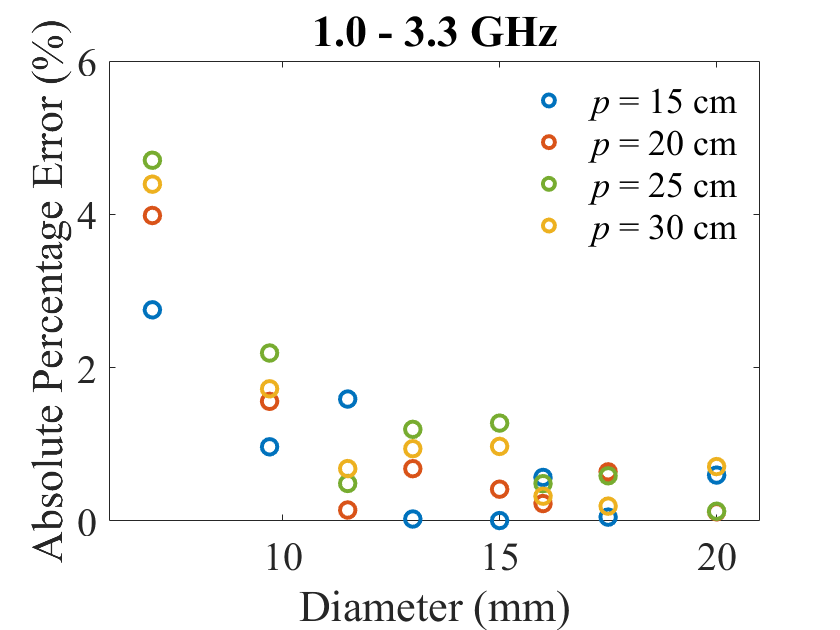}}\\
        \caption{Measurement results and estimation errors in the air. (a) Measured power ratios of metal bars of different diameters at different depths, and the calculated theoretical curves in three different frequency bands. (b) The estimation errors of the proposed method in different frequency bands.}
        \label{fig:air}
    \end{center}
\end{figure*}

\subsection{Measurement in the Air}
The experimental setup in the air is illustrated in Fig. \ref{fig:conf}. Eight cylindrical metal bars are used as samples. Their diameters measured by a caliper are 7 mm, 10 mm, 11.5 mm, 13 mm, 15 mm, 16 mm, 18 mm, and 20 mm, respectively. A foam box and an acrylic fixture are used to fix the position of the antenna and metal bar. The metal bar is placed right below the antenna at four depths (15 cm, 20 cm, 25 cm, and 30cm).  In the experiment, the control unit and the VNA are placed far away from the antenna to avoid their interference. For each metal bar at each depth, two A-scans of the two orthogonal polarizations are recorded in the frequency domain from 0.5 GHz to 3.3 GHz. To extract the scattered signal of the metal bar, background subtraction is performed by measuring a background A-scan in the same environment but without the metal bar and subtracting it from the A-scan with the metal bar. \par

Three frequency bands are used to analyze the data, which are 0.5 – 2.8 GHz, 0.5 – 3.3 GHz, and 1.0 – 3.3 GHz, respectively. The collected frequency domain data are transformed to the impulse response in the time domain via inverse Fourier transform to calculate the wideband power ratio. The measured power ratio of metal bars of different diameters at different depths is shown in Fig. \ref{air1}. The power ratios of the same metal bar at different depths have almost identical values, which verifies that the power ratio is independent of the depths in the far field. With the increment of the frequency spectrum, the power ratio is increased, which agrees with the influence of the frequency spectrum on the power ratio as discussed in Section \ref{sec3}.

The theoretical curves are calculated by \eqref{eq_10} in different frequency ranges. Same as shown in the simulated results, the theoretical curves in different depths overlap each other, so we plot the averaged curve in Fig. \ref{air1} and use it to estimate the diameter of the metal bar. As shown in Fig. \ref{air1}, the measured power ratios of the metal bar of different diameters are in good agreement with their corresponding theoretical values. Therefore, the estimated diameters based on the theoretical curves are very close to the real diameters. The estimation errors are plotted in Fig. \ref{air2}. The errors are all less than 6\% and a majority of errors are less than 3\%. The maximum error appears at the smallest diameter of 7 mm. This is because the theoretical curve around the 7 mm region has a small slope, so a small variation in the measured power ratio due to the imperfect background subtraction can result in a large deviation of the estimated diamater from the real diameter. The slope of the curve in the small diameter region increases with the increment of the frequency spectrum, leading to a decrease in error from 5.26\% to 4.71\%. Since the theoretical curve in the 1.0 - 3.3 GHz band has the largest slope for the given diameter range, the diameter estimated in this frequency band is the least sensitive to interference from residue environmental noise, and therefore the frequency band produces the most accurate results with the averaged error of 1.10\%.

\begin{figure*}[!t]
    \begin{center}
        \subfigure[]{
        \label{sand1}
        \includegraphics[width=0.33\linewidth]{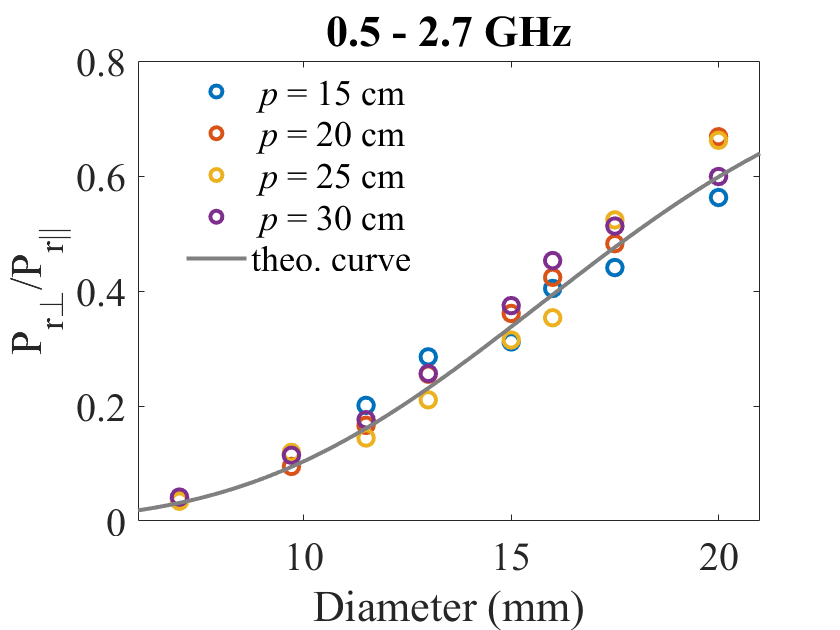} 
		\includegraphics[width=0.33\linewidth]{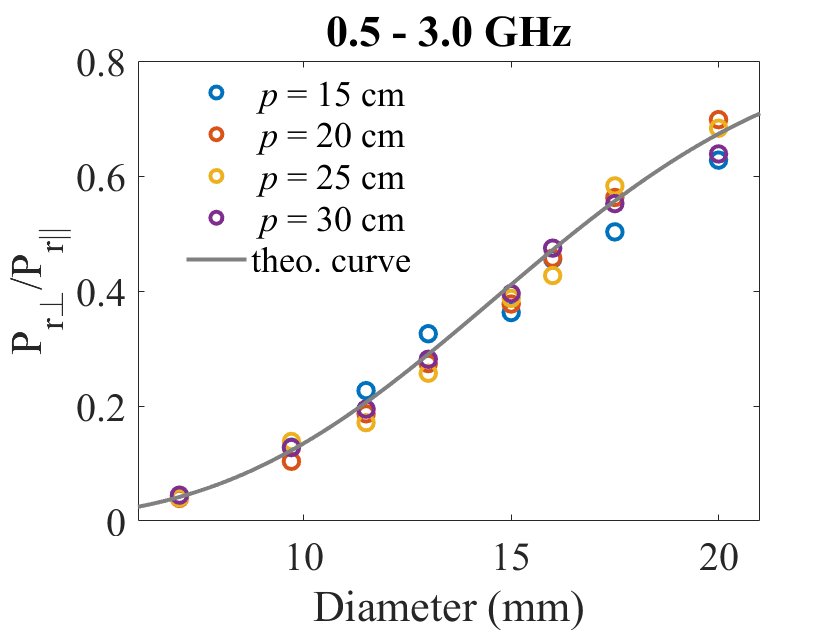} 
		\includegraphics[width=0.33\linewidth]{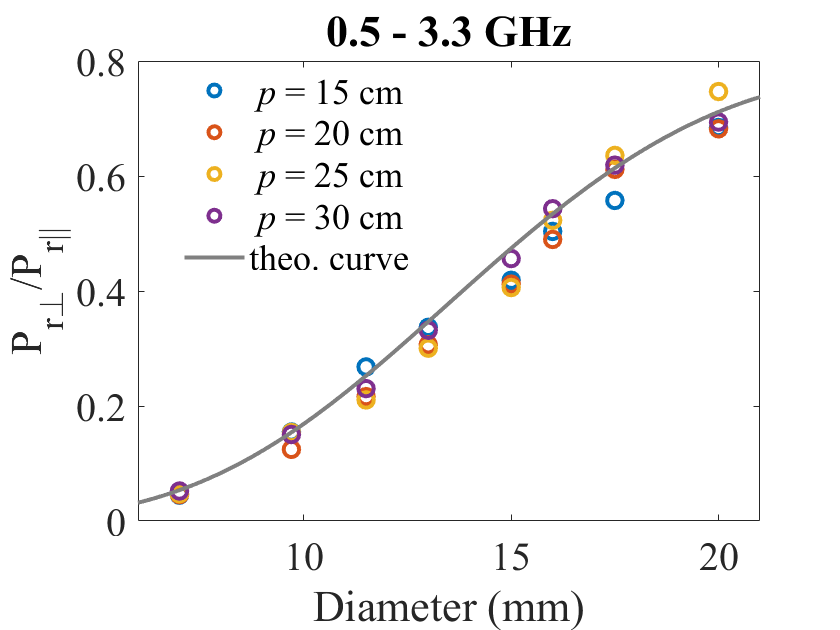}}\\
        \subfigure[]{
        \label{sand2}
        \includegraphics[width=0.33\linewidth]{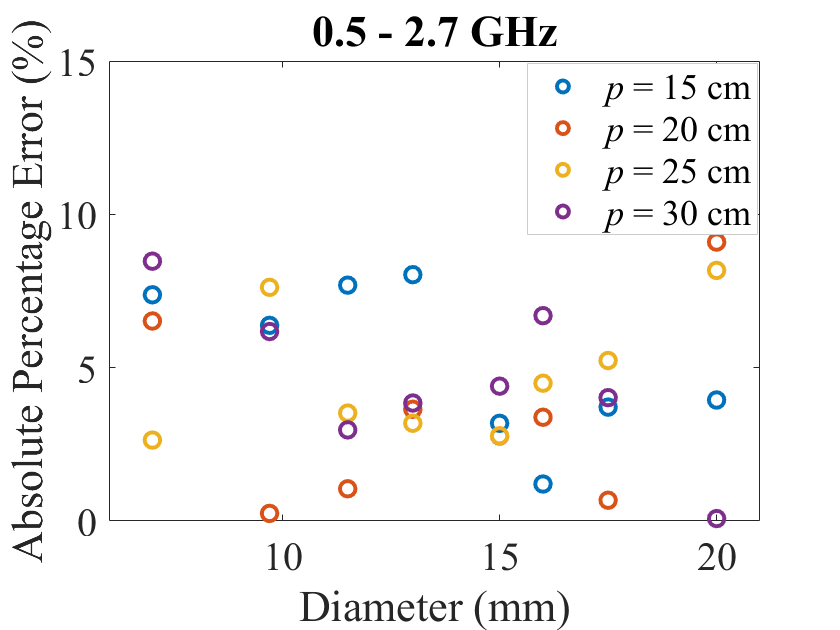} 
		\includegraphics[width=0.33\linewidth]{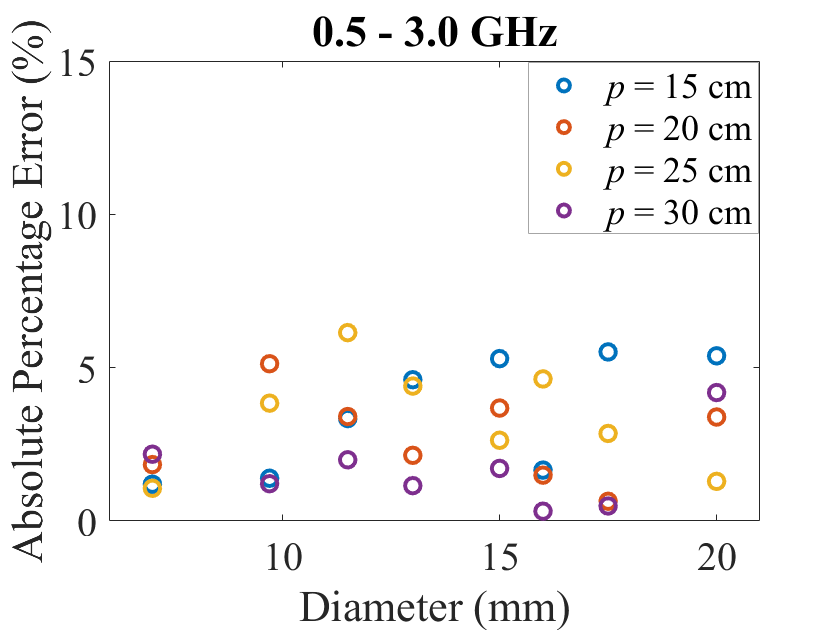} 
		\includegraphics[width=0.33\linewidth]{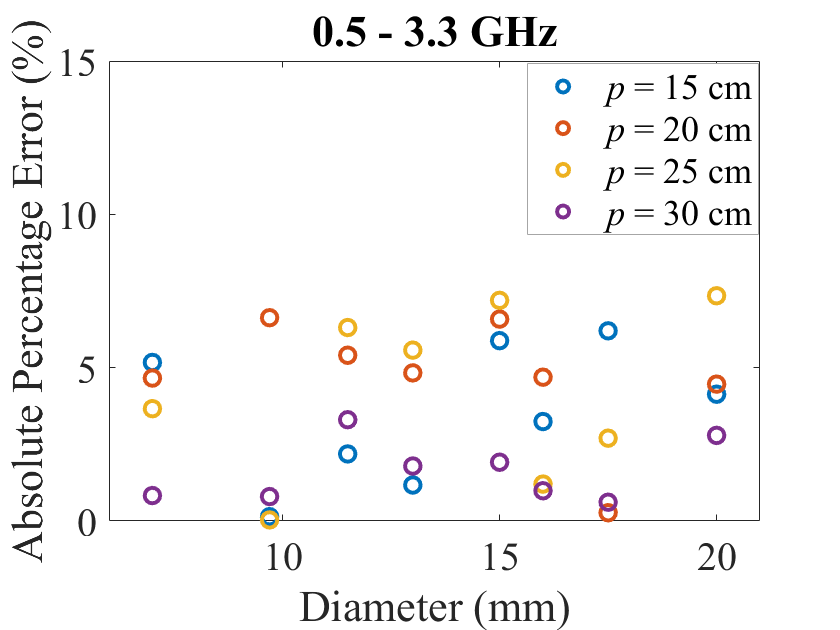}}\\
        \caption{Measurement results and estimation errors in the sand environment. (a) Measured power ratios of metal bars of different diameters at different depths, and the calculated theoretical curves in three different frequency bands. (b) The estimation errors of the proposed method in different frequency bands.}
        \label{fig:sand}
    \end{center}
\end{figure*}

\subsection{Measurement in Coarse Sand}
The sand environment is built by filling the foam box in Fig. \ref{fig:conf} with coarse sand. Same as the measurement in air, the eight metal bars of different diameters are placed at four different depths. The scattered waves of the metal bar at each depth are acquired by the dual-polarized GPR system in the frequency domain. Background removal is performed by subtracting a reference trace measured in the sand environment without the metal bar to extract the metal bar's response. Three different frequency bands are selected to analyze the data, which are 0.5 – 2.7 GHz, 0.5 – 3.0 GHz, and 0.5 – 3.3 GHz, respectively. The measured wideband power ratio of the metal bar at different depths in different bands is shown in Fig. \ref{sand1}. The power ratio of the same metal bar at different depths varies within a small range, which is due to the interference from background noise that cannot be fully eliminated using background subtraction in the coarse sand environment. \par

\begin{figure}[!h]
    \begin{center}
        \subfigure[]{
        \label{concrete1}
        \includegraphics[width=0.86\linewidth]{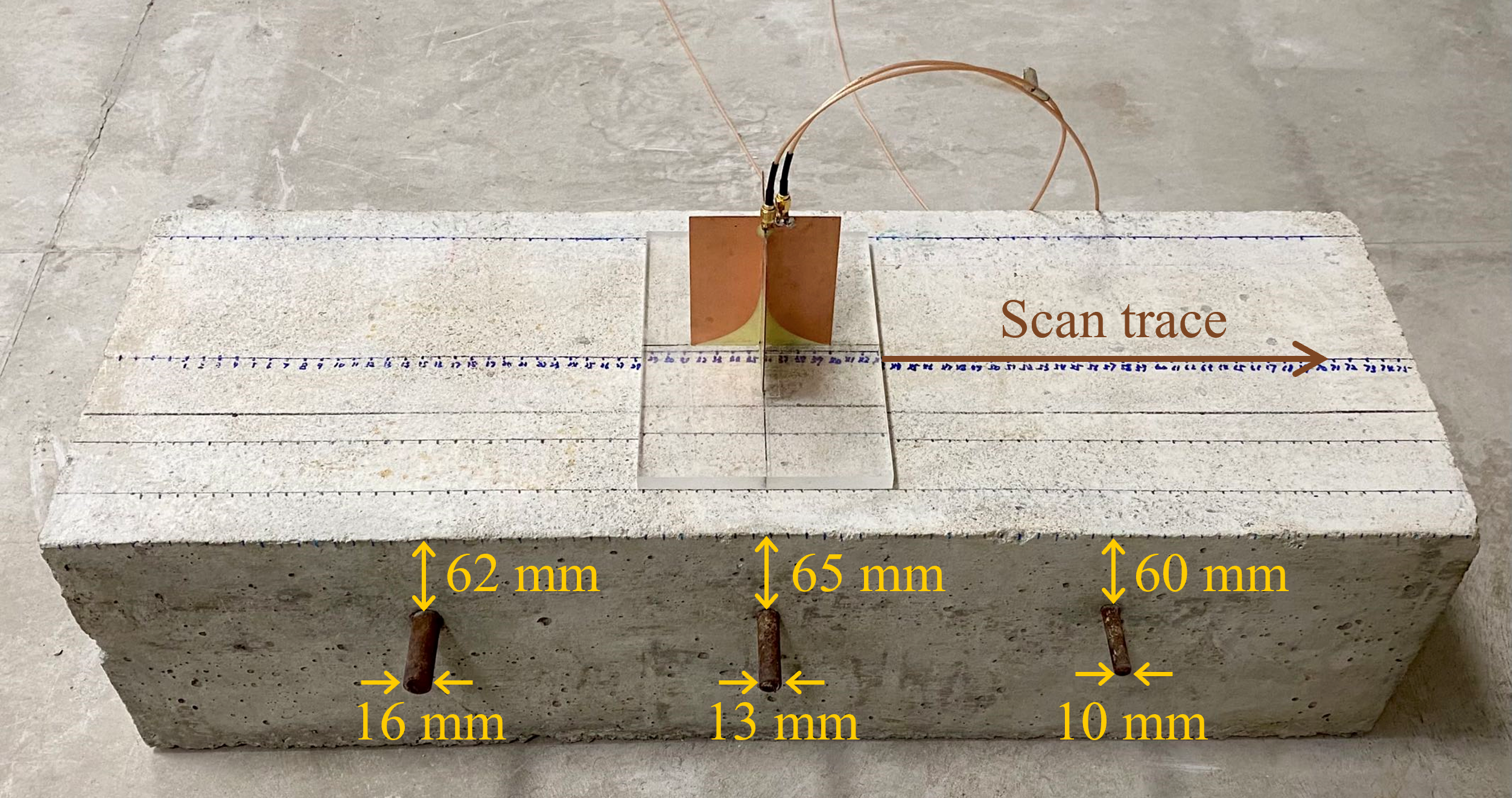}} \\
        \vspace{-0.1cm} 
        \subfigure[]{
        \label{bscan}
        \includegraphics[width=0.91\linewidth]{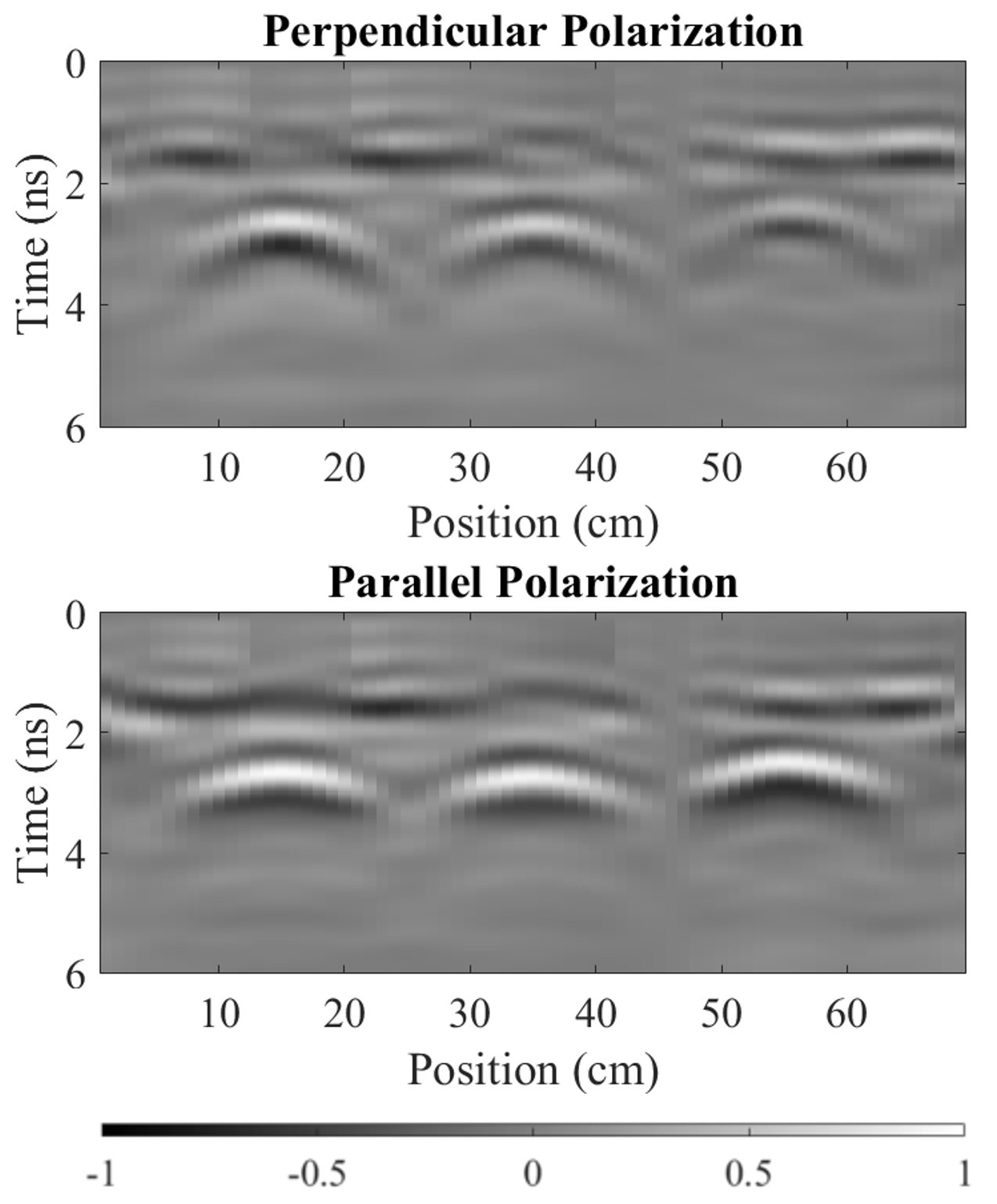}}\\
        \vspace{-0.1cm} 
        \subfigure[]{
        \label{result}
        \includegraphics[width=0.93\linewidth]{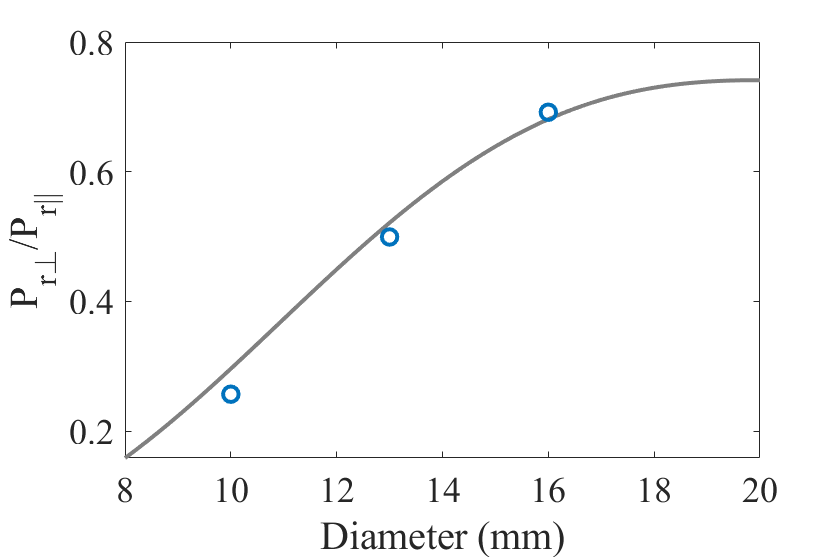}}\\
        \vspace{-0.1cm} 
        \caption{(a) The concrete specimen. (b) The measured B-scans of the concrete of the perpendicular polarization and the parallel polarization. (c) The measured power ratio and the theoretical curve. The diameters of the metal bars are accurately estimated based on the theoretical curve with a maximum error of 5.30\%.}
        \label{fig:concrete}
    \end{center}
\end{figure}

The calculation of the theoretical curve requires the pre-determined relative permittivity value of the sand. To obtain that, a metal plate is placed at a known depth $D$ in the sand and the delay time from the sand surface to the plate reflection $\Delta t$ is measured. The relative permittivity of the sand is then calculated by $(c\Delta t/2D)^2$. The measurement is conducted three times by placing the metal plate at different depths, and the averaged relative permittivity value of the sand is obtained as 2.66. The theoretical curves are then calculated by \eqref{eq_10} in different frequency ranges, which are also plotted in Fig. \ref{sand1}.  The measured power ratios are close to their corresponding theoretical values. The estimation error based on the power ratio and the theoretical curves are plotted in Fig. \ref{sand2}. The errors are all less than 10\%. The increase in error compared to the ideal air environment is because the background noise cannot be well subtracted in the heterogeneous sand environment, leaving residual noise that interferes with reflected signals of the metal bar. Compared with the estimation results in the 0.5 - 2.7 GHz band, increasing the frequency band yields a larger slope of the theoretical curves, thereby reducing the estimation sensitivity to the variation of the power ratio. With a larger slope of the theoretical curve in the 0.5 - 3.0 GHz and 0.5 - 3.3 GHz bands, the errors for the metal bar of different diameters are consistent and within 7.3\%. \par
The experiment results in the air and sand environments not only verify the effectiveness of the method in sizing metal bars at different depths in different mediums, more importantly, it also demonstrates that the frequency range of a wideband GPR can be tailored to produce a larger slope of the theoretical curve in the given medium so as to achieve more accurate sizing performance. \par

\subsection{Measurement in Concrete} \label{sec4.3}

The third experiment is conducted on a concrete sample with three embedded metal bars, as shown in Fig. \ref{concrete1}. The three metal bars have diameters of 16 mm, 13 mm, and 10 mm, and are located at cover depths of 62 mm, 65 mm, and 60 mm, respectively. The bar spacing is 20 cm. B-scans of the perpendicular and parallel polarizations are collected by moving the dual-polarized antenna along a scanning trace that is perpendicular to the metal bars' axes on the concrete surface. The frequency band 0.5 GHz - 2.5 GHz is used to process the data as this frequency band guarantees a proper slope of the theoretical curve to estimate the metal bar size, which will be shown in the result. To remove background noise, an A-scan acquired between the two smallest metal bars using the perpendicular polarization is used as the background trace and subtracted from all traces in the B-scans. The B-scans of the two polarized signals after background removal are shown in Fig. \ref{bscan}. The maximum amplitudes of the reflected signals of the metal bars are extracted from the B-scan, and the power ratios are calculated and plotted in Fig. \ref{result}. \par

To obtain the relative permittivity of the concrete, a metal plate is placed at the bottom of the concrete and the delay time from the top and bottom surface $\Delta t$ is measured. The relative permittivity of the concrete is calculated using the height of the concrete $H$ and $\Delta t$ by $(c\Delta t/2H)^2$. The relative permittivity values of three different locations are measured, and their average value of 8.0 is used in the calculation of the theoretical curve of the power ratio. The three parallel polarized A-scan traces acquired on top of the three metal bars are used to calculate the frequency spectrum of the parallel polarization. Three theoretical curves are calculated using the three frequency spectra and they overlap with each other. Their averaged values are used as the final theoretical curve, as plotted in Fig. \ref{result}. Based on the measured power ratio and the theoretical curve shown in Fig. \ref{result}, the diameters of the three metal bars are estimated as 16.32 mm, 12.68 mm, and 9.47 mm, respectively. The estimated values are very close to the real values with a maximum percentage error of 5.30\%. The estimation error can be caused by the interference of the residue background noise. Nevertheless, the results have shown promising accuracy in estimating the size of the metal bar in concrete.

\section{Discussion and Comparison}\label{sec5}

\subsection{Estimation Error Caused by Inaccurate Relative Permittivity}

\begin{figure}[!h]
    \begin{center}
        \subfigure[]{
        \label{eps_error1}
        \includegraphics[width=0.91\linewidth]{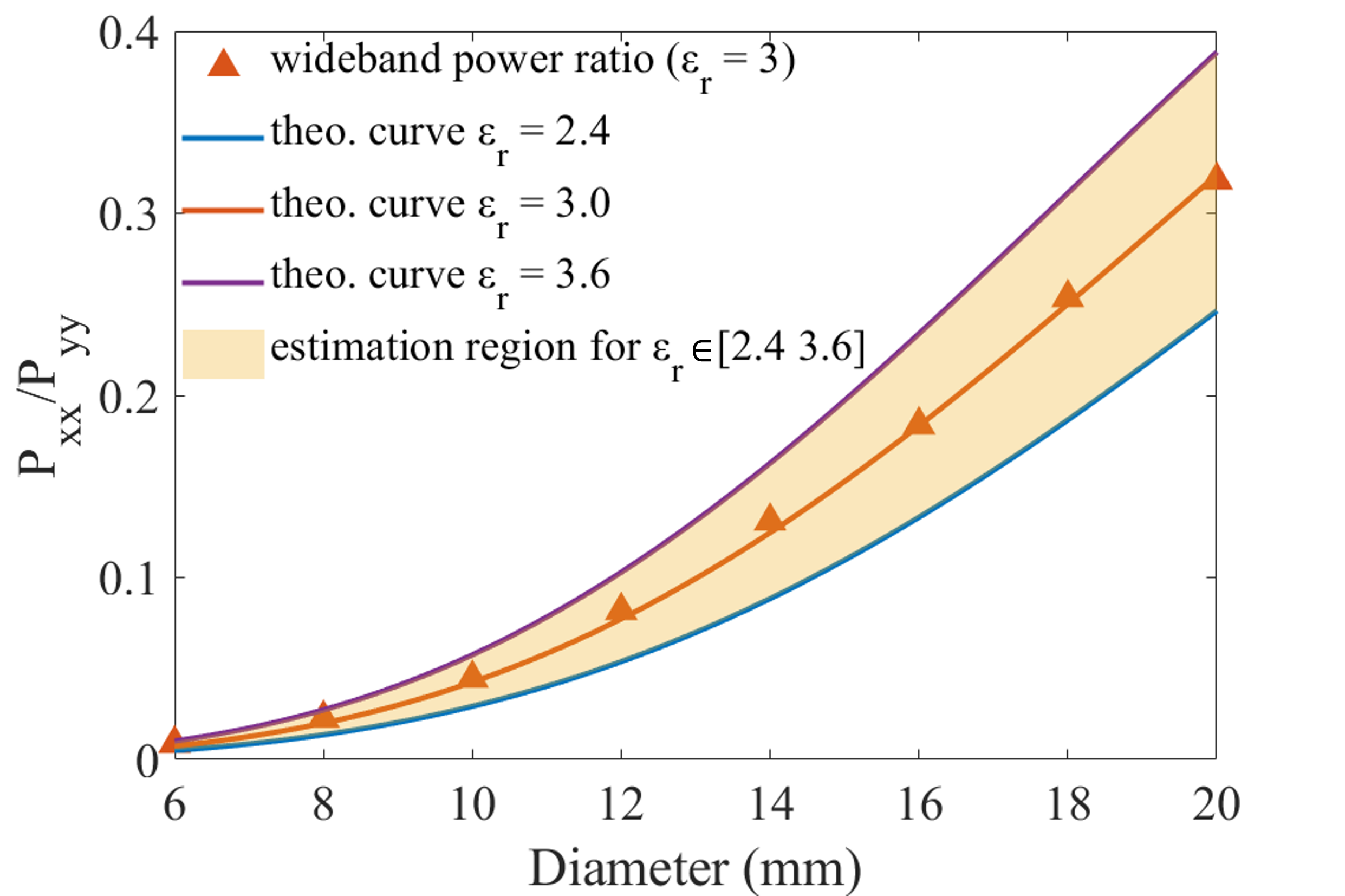}} \\
        \vspace{-0.1cm} 
        \subfigure[]{
        \label{eps_error2}
        \includegraphics[width=0.91\linewidth]{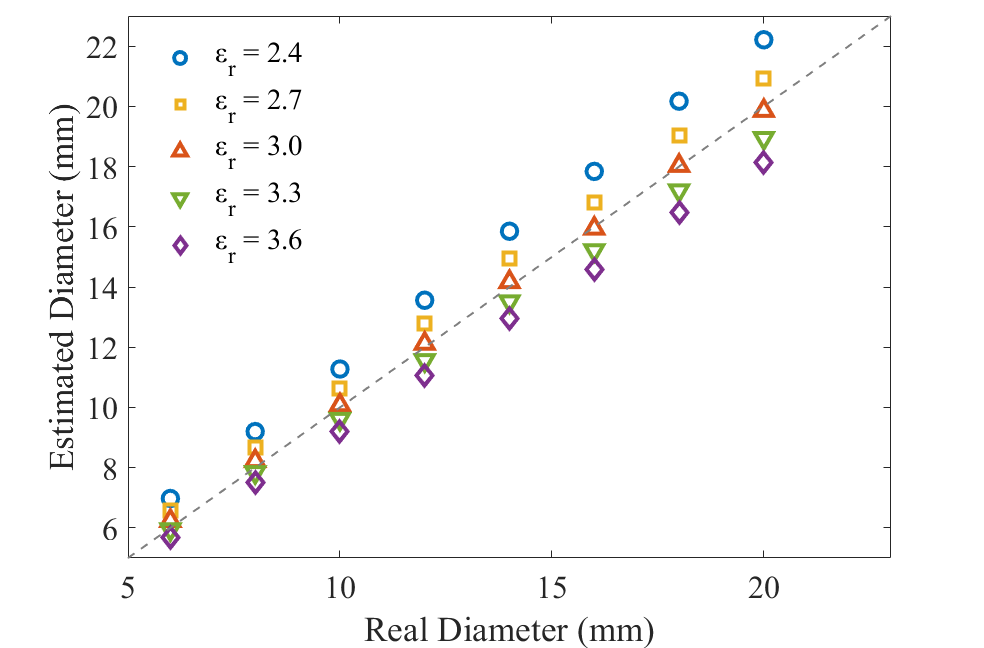}}\\
        \vspace{-0.1cm} 
        \subfigure[]{
        \label{eps_error3}
        \includegraphics[width=0.91\linewidth]{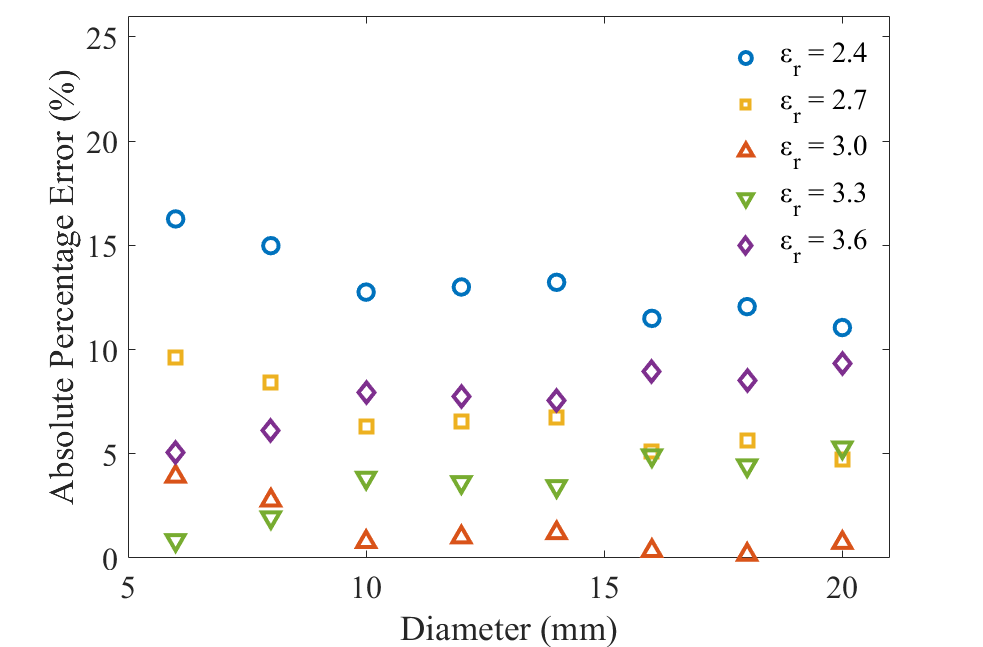}}\\
        \vspace{-0.1cm} 
        \caption{(a) The wideband power ratios of a metal bar of different diameters when $\epsilon_r$=3 compared with theoretical curves computed using different $\epsilon_r$. (b) The estimated diameters, and (c) the absolute percentage errors using theoretical curves with different $\epsilon_r$.}
        \label{fig:eps_error}
    \end{center}
    \vspace{-0.5cm}
\end{figure}

As the proposed method requires prior knowledge of relative permittivity ($\epsilon_r$) of the subsurface medium, the accuracy of $\epsilon_r$ affects the estimation accuracy of the bar diameter $d$.  A case study is performed to demonstrate the influence of $\epsilon_r$ on the estimation accuracy. In the case study, a GPR with 1-GHz Ricker waveform is used to detect a metal bar with different diameters in a medium with $\epsilon_r$ of 3. The wideband power ratios of metal bars of different diameters are shown in Fig.  \ref{eps_error1}. The theoretical curves obtained using the proposed method with $\epsilon_r$ ranging from 2.4 to 3.6 are also shown in Fig. \ref{eps_error1}. The estimated diameter values and the estimation errors using curves with different $\epsilon_r$ are shown in Fig. \ref{eps_error2} and Fig. \ref{eps_error3}, respectively. It can be seen from Fig. \ref{eps_error2} that smaller $\epsilon_r$ tends to overestimate the diameter, whereas larger $\epsilon_r$ underestimates the diameter. As shown in Fig. \ref{eps_error3}, the estimation errors are maintained within 10\% when $\epsilon_r$ deviates within 10\% from the actual value ($\epsilon_r\in$ [2.7, 3.3]). A similar phenomenon has been verified using different GPR operating frequencies in our study. Therefore, a slight inaccuracy of $\epsilon_r$ within 10\% would not cause a large estimation error of the proposed method. 

In addition, as shown in Fig. \ref{eps_error1}, the curve corresponding to the correct $\epsilon_r$ value can also be determined based on a bar with known diameter. Therefore, the proposed method can also be calibrated based on a bar with known diameter as an alternative to $\epsilon_r$ measurement.

\subsection{Effective Diameter Estimation Range of the Method}

\begin{figure}[!t]
    \begin{center}
        \includegraphics[width=0.93\linewidth]{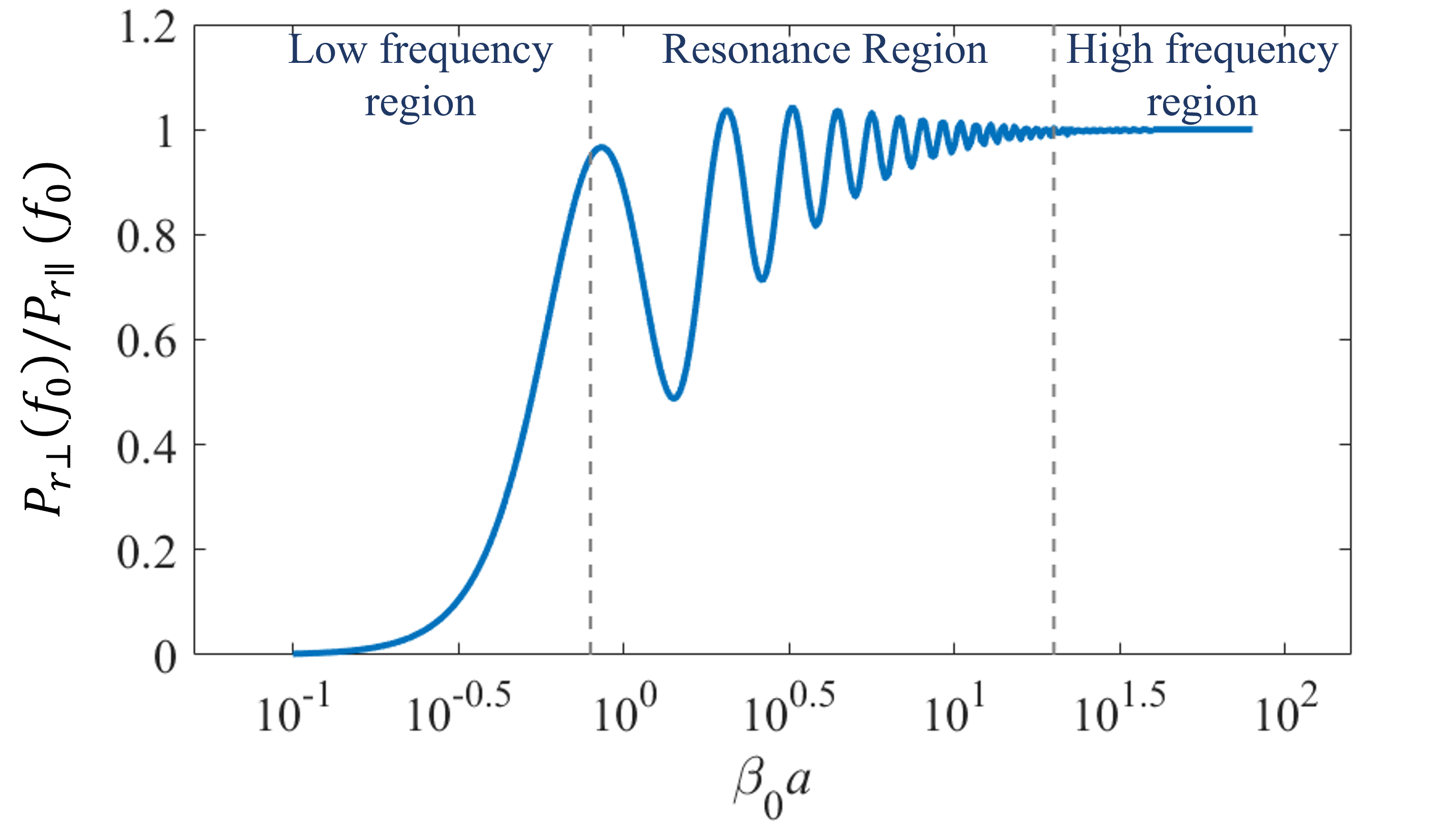}\\
        \caption{The power ratio at a single frequency as a function of $\beta_0 a$.}
        \label{fig:singlefreq_limit}
    \end{center}
    \vspace{-0.5cm}
\end{figure}

\begin{figure}[!t]
    \begin{center}
        \subfigure[]{
        \label{wideband_freq}
        \includegraphics[width=0.48\linewidth]{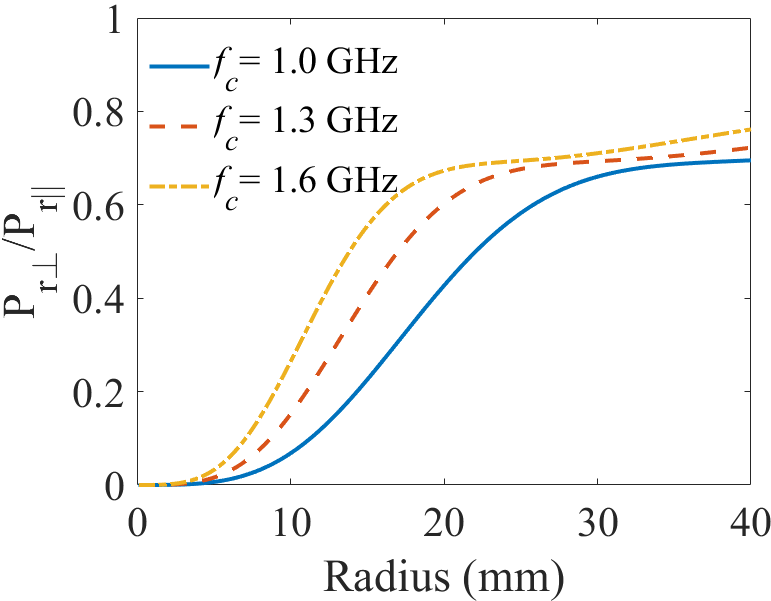}}
        \subfigure[]{
        \label{wideband_permittivity}
        \includegraphics[width=0.48\linewidth]{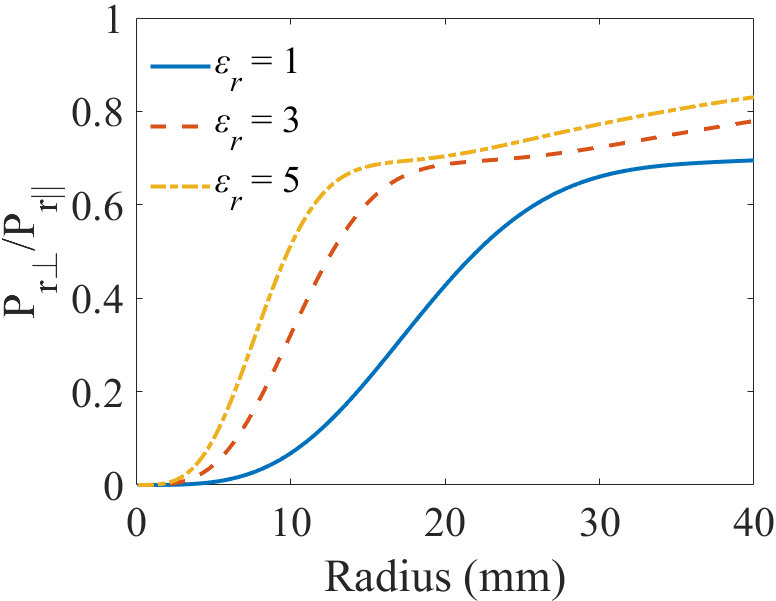}}
        \caption{The wideband power ratio as a function of bar radius in the cases of (a) GPR spectra with different center frequency $f_c$ when $\epsilon_r$ = 1, and (b) subsurface medium with different relative permittivity $\epsilon_r$ when $f_c$ = 1 GHz.}
        \label{fig:wideband_limit}
    \end{center}
    \vspace{-0.5cm}
\end{figure}

Since the proposed method is based on the relationship between the power ratio and the bar radius, it is only effective when the relationship is unique. To find the effective diameter range that can guarantee the unique relationship, the relationship between the single-frequency power ratio and the bar radius, and the relationship between the wideband power ratio in \eqref{eq_10} and the bar radius are both investigated.

The power ratio at a single frequency $f_0$  is calculated by \eqref{eq_2}-\eqref{eq_5}. The relationship between the single-frequency power ratio and $\beta_0 a$ is plotted in Fig. \ref{fig:singlefreq_limit}. As shown in Fig. \ref{fig:singlefreq_limit}, the relationship between the power ratio and $\beta_0 a$ is divided into three regions: the low frequency region, the resonance region, and the high frequency region \cite{radarbook}. Only the low frequency region ($\beta_0 a\ll$1) guarantees a monotonic increase in power ratio with $\beta_0 a$, which can be used for unambiguous radius estimation. 

Since \eqref{eq_10} is the extension of the single-frequency power ratio  in a wideband form, a similar relationship exists between the wideband power ratio and the radius. Fig. \ref{fig:wideband_limit} shows the wideband power ratio as a function of bar radius $a$ in the cases of GPR spectra with different center frequencies $f_c$ and subsurface medium with different subsurface permittivity $\epsilon_r$. We experimentally found that the curves keep their monotonicity when $a<\frac{30}{f_c \sqrt{\varepsilon_r}}$, where $a$ is in mm and $f_c$ is in GHz. Therefore, to ensure unambiguous radius estimation for a metal bar with a large radius in a medium with high relative permittivity, a GPR with a lower center frequency is needed. The method is particularly suitable for estimating small-radius bars, which complements the mainstream reflection pattern-based GPR methods  that are more effective for large-radius bars ($a>$10 mm).

\subsection{Diameter Measurement in Different Construction Configurations}
Experimental results in Section \ref{sec3} and Section \ref{sec4} have demonstrated the performance of the proposed method in single metal bar cases, whereas in real scenarios, there are multiple metal bars arranged in different configurations. In this subsection, we demonstrate that the proposed method maintains its effectiveness as long as the reflection of the target bar is separated from those of other bars around it.

\begin{figure}[t]
    \begin{center}
        \includegraphics[width=0.98\linewidth]{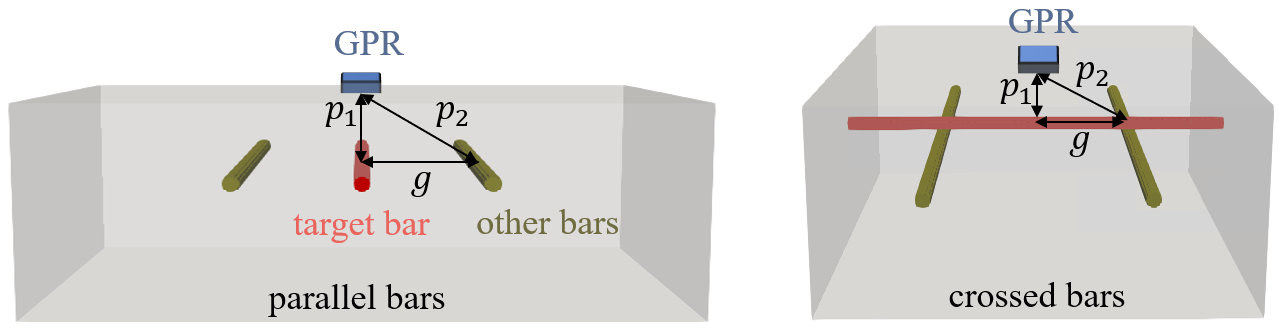}
        \caption{A scenario of multiple metal bars.}
        \label{fig:illustration_bardistance}
    \end{center}
    \vspace{-0.5cm}
\end{figure}

As shown in Fig. \ref{fig:illustration_bardistance}, to guarantee the sufficient separation of reflections of the target bar and other bars when the GPR is located above the target bar for detection, the distance from these metal bars to the GPR ($p_1$ and $p_2$) based on the GPR resolution should satisfy
\begin{equation}
\label{eq_11}
\left|p_2-p_1\right| \geq \frac{c}{2 f_c \sqrt{\varepsilon_r}},
\end{equation}
If the metal bars are located at the same depth as shown in Fig. \ref{fig:illustration_bardistance}, the distance $g$ between the closest point of the target bar to the GPR antenna and that of the other bars to the GPR antenna should satisfy 
\begin{equation}
\label{eq_12}
\left(\sqrt{g^2+p_1^2}-p_1\right) \geq \frac{c}{2 f_c \sqrt{\varepsilon_r}}.
\end{equation}

\begin{figure*}[!t]
    \begin{center}
        \includegraphics[width=0.9\linewidth]{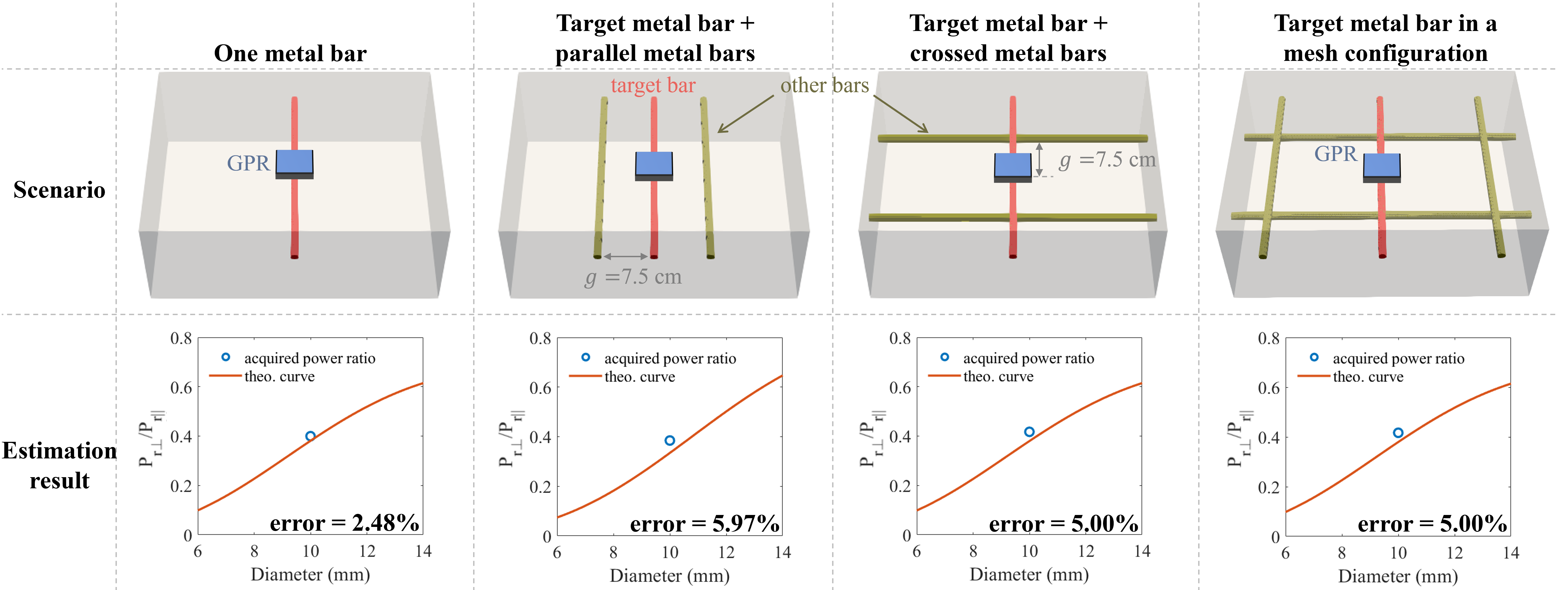}
        \caption{Illustrations and estimation results of four scenarios: i) one metal bar, ii) the target metal bar and parallel metal bars, iii) the target metal bar and cross metal bars, and iv) the target metal bar in a mesh configuration.}
        \label{fig:multiplebar}
    \end{center}
    \vspace{-0.5cm}
\end{figure*}

\begin{figure}[t]
    \begin{center}
        \includegraphics[width=0.9\linewidth]{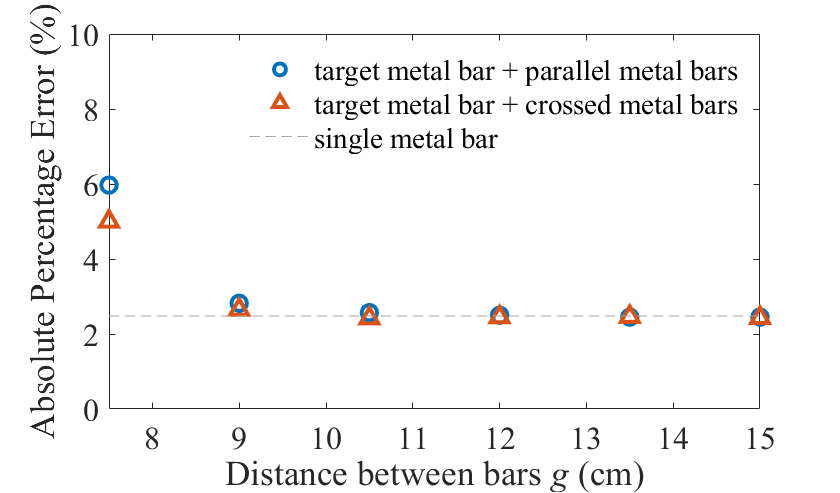}
        \caption{The estimation error of the target metal bar’s diameter with different distances between metal bars $g$.}
        \label{fig:error_bardistance}
    \end{center}
    \vspace{-0.5cm}
\end{figure}

Four scenarios are simulated to demonstrate the performance of the proposed method in different construction configurations satisfying the distance constraint in \eqref{eq_11}. The scenarios are i) one metal bar, ii) the target metal bar (as marked in red) with parallel metal bars, and iii) the target metal bar and crossed metal bars, and iv) the target metal bar in a mesh configuration, as shown in Fig. \ref{fig:multiplebar}. The metal bars have a diameter of 10 mm and are located at a depth of 5 cm in concrete with relative permittivity of 8. A dual-polarized GPR with a 1.3-GHz ricker waveform is used to acquire the reflections of the metal bars. As shown in Fig. \ref{fig:multiplebar}, the bars in the parallel bar case are separated by the minimum distance $g$ = 7.5 cm as calculated using \eqref{eq_12}. In the crossed bar case, the distance between the crossed bars is 2$g$ = 15.0 cm and the GPR is positioned in the middle between the two crossed bars to measure the target bar. The measured wideband power ratio of the target metal bar and the calculated theoretical curve in these cases are shown in Fig. \ref{fig:multiplebar}. Since the reflection of the target metal bar is separated from those of other metal bars, the wideband power ratio can be reliably extracted and used for diameter estimation. The estimation errors for the four scenarios based on the proposed method are 2.48\%, 5.97\%, 5.00\% and 5.00\%, respectively. Compared with the estimation error in the single bar case, the errors in other construction configurations with multiple neighboring bars are only slightly increased. If the bar distance is further increased, the estimation error will decrease to the level of that of a single bar, as shown in Fig. \ref{fig:error_bardistance}. The results demonstrate that the method maintains high accuracy in different construction configurations as long as the distance between adjacent bars satisfies the constraint in \eqref{eq_11}.

\subsection{Diameter Measurement of a Metal Bar of Different Orientations}

Since the proposed method is developed based on the scattering width of cylindrical metal bars at parallel and perpendicular linear polarizations, orthogonally linear polarized GPRs have been used in Sections \ref{sec3} and \ref{sec4} as the most straightforward way to demonstrate the effectiveness of the method. The current application of the method requires prior knowledge of the bar orientation to align the polarization direction. 

To further extend the applicability of the proposed method to different bar orientation cases, quad-circularly-polarized systems that acquire the full scattering matrix $\left[\begin{array}{ll}
S_{L L} & S_{R L} \\
S_{L R} & S_{R R}
\end{array}\right]$ and quad-linearly-polarized systems that acquire $ \left[\begin{array}{ll}
S_{V V} & S_{H V} \\
S_{V H} & S_{H H}
\end{array}\right]$ can be used. The scattering matrices of the circular and linear polarization can also be transformed into each other using 
\begin{equation}
\label{eq_13}
\left[\begin{array}{ll}
S_{V V} & S_{H V} \\
S_{V H} & S_{H H}
\end{array}\right]=\frac{1}{2}\left[\begin{array}{cc}
1 & 1 \\
-j & j
\end{array}\right]\left[\begin{array}{ll}
S_{L L} & S_{R L} \\
S_{L R} & S_{R R}
\end{array}\right]\left[\begin{array}{cc}
1 & -j \\
1 & j
\end{array}\right].
\end{equation}

Based on the scattering matrix, the bar orientation angle  $\theta$ can be extracted using several well-developed methods in \cite{pol4,rotation1}. Once the orientation angle $\theta$ is obtained, one can either replan the scanning trace, or directly transform the measured data to the polarizations that are parallel and perpendicular to the metal bar using \eqref{eq_14} \cite{rotation1}. $S_{\|}$ and $S_{\perp}$ can then be used to implement the proposed method.

\begin{figure*}[b]
\begin{equation}
\label{eq_14}
\left[\begin{array}{cc}
S_{\|} & S_{\perp \|} \\
S_{\| \perp} & S_{\perp}
\end{array}\right]=\left[\begin{array}{cc}
\cos \theta & \sin \theta \\
-\sin \theta & \cos \theta
\end{array}\right]\left[\begin{array}{ll}
S_{H H} & S_{H V} \\
S_{V H} & S_{V V}
\end{array}\right]\left[\begin{array}{cc}
\cos \theta & \sin \theta \\
-\sin \theta & \cos \theta
\end{array}\right]^T.
\end{equation}
\end{figure*}

\subsection{Comparison with Existing Methods}

The estimation accuracy of the proposed method is compared with that of existing GPR methods in the concrete case with three embedded metal bars as described in Section \ref{sec4.3}. The GPR methods compared include the reflection pattern-based method \cite{curvefittinglimit} and the power ratio method at the GPR nominal frequency \cite{DP3}. Table \ref{table:comp} shows the estimation results of different methods. The reflection pattern-based method produces large errors for these small-diameter metal bars, which is consistent with the results in \cite{curvefittinglimit}. This is because the limited resolution of the GPR cannot produce noticeable differences in the hyperbolic curvature with small variations in diameter, and a small amount of environmental clutter makes the fitting problem ill-posed \cite{curvefittinglimit}. The power ratio method at the GPR nominal frequency produces relatively large estimation errors in the 13-mm and 16-mm metal bar cases. The discrepancy is because the GPR is a wideband device whose performance cannot be accurately characterized by the single nominal frequency point. The proposed method as a reflection power-based method circumvents the resolution limit of the pattern-based methods and addresses the single-frequency limit in \cite{DP3} by taking both the wideband GPR spectrum and scattering width of metal bars into account. It achieves the highest accuracy in all cases. The comparison results demonstrate the effectiveness of our proposed method in improving the accuracy of bar diameter estimation.

\begin{table*}
\caption{Comparison of Estimation Accuracy of Different GPR Methods}
\centering
\arrayrulecolor{black}
\begin{tabular}{l|llllll} 
\hline\hline
\multicolumn{1}{c|}{\multirow{2}{*}{\diagbox{Method}{Diameter (\textit{d})}}} & \multicolumn{2}{c}{10 mm}                                & \multicolumn{2}{c}{13 mm}                                 & \multicolumn{2}{c}{16 mm}                                  \\ 
\cline{2-7}
\multicolumn{1}{c|}{}                                                         & $d_{est}$                   & Error                      & $d_{est}$                   & Error                      & $d_{est}$                   & Error                       \\ 
\hline
Reflection pattern-based method \cite{curvefittinglimit}                                        & 15.62 mm                    & 56.20\%                    & 16.80 mm                     & 29.23\%                    & 20.59 mm                     & 28.70\%                     \\ 
\hline
Power ratio method at GPR nominal frequency \cite{DP3}                            & 9.32 mm                     & 6.80\%                     & 11.60 mm                     & 10.77\%                    & 13.69 mm                     & 14.44\%                     \\ 
\hline
Our method                                                                    &  9.47 mm  &  5.30\%  & 12.68 mm  &  2.46\%  &  16.32 mm  &  2.00\%  \\
\hline\hline
\end{tabular}
\arrayrulecolor{black}
\label{table:comp}
\end{table*}

\section{Conclusion}\label{sec6}
This paper presents a novel measurement and data analysis method that uses a wideband dual-polarized GPR to estimate the diameter of subsurface metal bars. The theoretical relationship between the bar diameter and the wideband power ratio of the bar reflected signals acquired by the dual-polarized GPR is established via the scattering width of the metal bar and the spectrum of the bar reflected signal. Based on the theoretical model, the bar diameter can be reliably estimated using the obtained power ratio in a GPR survey. The effectiveness of the method has been verified using simulated and measured data collected by GPRs with different frequency spectra in different mediums. Experimental results show that the method can accurately estimate the diameter of metal bars with an absolute percentage error of less than 10\%. As a reflection power-based method, it circumvents the resolution limit of the conventional reflection-pattern based methods and outperforms the conventional methods by a large margin. The high accuracy and simplicity of the proposed method make it suitable for the on-site subsurface metal bar measurement.   \par

\end{document}